\documentclass[aps,prb,twocolumn,superscriptaddress,floatfix]{revtex4}
\usepackage{amsmath,amssymb,bm}
\usepackage{graphicx}
\usepackage{epstopdf}
\usepackage{latexsym}
\usepackage{subfigure}
\usepackage{color}

\newcommand{\ccc}{{Cs$_2$CuCl$_4$}}
\newcommand{\ccb}{{Cs$_2$CuBr$_4$}}

\def\DM{\text{Dzyaloshinskii-Moriya}}
\def\be{\begin{equation}}
\def\ee{\end{equation}}
\def\bea{\begin{eqnarray}}
\def\eea{\end{eqnarray}}

\begin{document}

\title{Deformed triangular lattice antiferromagnets in a magnetic field: role of spatial anisotropy and 
\DM~interactions}

\author{Christian Griset}
\affiliation{Department of Physics, California Institute of Technology, Pasadena, CA 91125}
\affiliation{Department of Physics, University of California, Santa Barbara, CA 93106}
\author{Shane Head}
\affiliation{Department of Physics and Astronomy, University of Utah, Salt Lake City, UT 84112}
\author{Jason Alicea}
\affiliation{Department of Physics and Astronomy, University of California, Irvine, California 92697}
\author{Oleg A. Starykh}
\affiliation{Department of Physics and Astronomy, University of Utah, Salt Lake City, UT 84112}

\date{July 6, 2011}

\begin{abstract}
  Recent experiments on the anisotropic spin-1/2 triangular antiferromagnet \ccb~have revealed a remarkably rich phase diagram in applied magnetic fields, consisting of an unexpectedly large number of ordered phases.  Motivated by this finding, we study the role of three ingredients---spatial anisotropy, \DM~interactions, and quantum fluctuations---on the magnetization process of a triangular antiferromagnet, coming from the semiclassical limit.  The richness of the problem stems from two key facts: 1) the classical isotropic model with a magnetic field exhibits a large accidental ground state degeneracy, and 2) these three ingredients compete with one another and split this degeneracy in opposing ways.  Using a variety of complementary approaches, including extensive Monte Carlo numerics, spin-wave theory, and an analysis of Bose-Einstein condensation of magnons at high fields, we find that their interplay gives rise to a complex phase diagram consisting of numerous incommensurate and commensurate phases.  Our results shed light on the observed phase diagram for \ccb~and suggest a number of future theoretical and experimental directions that will be useful for obtaining a complete understanding of this material's interesting phenomenology.  
\end{abstract}
\maketitle

\section{Introduction}
\label{sec:intro}

Antiferromagnetic spin models on the triangular lattice constitute one of the simplest and most widely studied realizations of geometric frustration.
Indeed, the Ising triangular antiferromagnet was the first spin model found to possess
a disordered ground state and extensive residual entropy\cite{wannier1950} at zero temperature ($T$). 
While the classical Heisenberg model on the triangular lattice does order at $T=0$ into a well-known
commensurate spiral pattern (also known as a $\sqrt{3}\times \sqrt{3}$ state), the fate of the quantum spin-1/2 Heisenberg Hamiltonian has been the subject of a long and fruitful debate spanning over 30 years of research.
Although the originally proposed {\em resonating valence bond liquid} \cite{anderson1973,fazekas1974}
did not emerge as the ground state of the spin-1/2 Heisenberg model \cite{huse1988,bernu1992,capriotti1999}, such a phase was later found in a related quantum dimer model
on the triangular lattice \cite{moessner2001}.

Triangular antiferromagnets in an applied magnetic field---which is our focus here---have also been extensively studied for decades, and found to possess unusual magnetization physics which remains only partially understood.  Underlying much of this interesting behavior is the discovery, made long ago \cite{kawamura1984}, that in a magnetic field
Heisenberg spins with isotropic exchange interactions exhibit a large {\em accidental classical ground state degeneracy}.  That is, at finite magnetic fields there exists an infinite number of continuously deformable classical spin configurations which constitute minimum energy states, but are in no way symmetry related.
As reviewed later, this degeneracy is lifted by thermal (finite $T$) \cite{kawamura1984}
and quantum (finite spin $S$) \cite{ChubukovGolosov} fluctuations;  
such fluctuation-driven selection, also known as order-by-disorder\cite{Shender,Henley},
results in a nontrivial temperature-field phase diagram \cite{kawamura1984} consisting of three ordered phases. 
In particular, coplanar `Y' and `V' states are separated by a collinear up-up-down (UUD)
phase which realizes, for finite spin $S$ and at $T=0$, a
one-third magnetization plateau over a finite field interval.  
This plateau state preserves 
continuous spin rotation symmetry about the magnetic field direction
and is remarkably stable: unlike {\em all} other magnetically-ordered
states it survives ``dimensional reduction'' and exists even in the
smallest possible triangular lattice strip, the two-chain zig-zag ladder \cite{okunishi2003,hikihara2010}.

To date a large number of magnetic materials have been synthesized that realize triangular antiferromagnets, and experiments on such compounds have highlighted the spectacular breadth of phenomena that can be driven by the interplay between magnetic fields and geometric frustration stemming from the lattice.  
For example, a stacked triangular antiferromagnet with weak inter-plane coupling is realized by the $S=5/2$ material RbFe(MoO$_4$)$_2$, whose phase
diagram\cite{svistov2006} features all of the fluctuation-selected states discussed above.
Reducing the magnitude of the magnetic ion's spin enhances quantum fluctuations, 
sometimes leading to highly non-classical behavior.  
Low-spin materials which constitute `deformed' triangular antiferromagnets---that is, with spatially anisotropic exchange interactions that are not SU(2) symmetric---have indeed provided numerous surprises which are very likely quantum in origin.  Studies of the spin-1/2 compounds \ccc~and \ccb~have been particularly fruitful in this regard.  
Inelastic neutron scattering has revealed striking dominance of a multi-particle continuum in the dynamical response 
of \ccc\ \cite{coldea2002,coldea2003}. This continuum is naturally explained in terms of 
spin-1/2 spinon excitations of weakly coupled chains. Ordered phases of this material show strong
sensitivity to the magnitude and direction of the external magnetic field \cite{tokiwa2006}. 
The complexity of the $h-T$
phase diagram has been attributed to the competition between several asymmetric exchanges of the
Dzyaloshinskii-Moriya (DM) type and a weak inter-plane exchange interaction. In particular it has been suggested \cite{starykh10}
that, although weak, inter-plane exchange can dominate over stronger but frustrated inter-chain coupling 
and dictate the type of three-dimensional magnetic ordering at low temperatures.

Our study here is strongly motivated by the isostructural material \ccb, whose triangular planes are less anisotropic and exhibit weaker inter-plane coupling compared to \ccc, leading to rather different but equally rich phenomenology in magnetic fields.  With fields directed in the plane of the triangular layers, the experimentally determined phase diagram for \ccb~hosts as many as \emph{nine phases} \cite{Fortune} at low temperature.  A one-third magnetization plateau features very prominently \cite{ono2003,tsujii2007,fujii2007} amongst 
the other, less understood, phases and offers a convenient starting point for theoretical
analysis \cite{UUDpaper}.  Notably, \ccc~shows no signs of this plateau, which has been attributed  \cite{starykh10} to its more pronounced spatial anisotropy and stronger inter-layer coupling\cite{ono2005,valenti2011}, which is known to suppress the plateau \cite{gekht1997}.  
And unlike the higher-spin material RbFe(MoO$_4$)$_2$ discussed above \cite{svistov2006},
the width of the magnetization plateau in \ccb~is essentially $T$-independent \cite{tsujii2007}, 
strongly hinting at its quantum origin.\cite{ChubukovGolosov,UUDpaper,tay2010}  

The sheer number of states present in the phase diagram of \ccb, in comparison with the three phases expected from the standard isotropic Heisenberg antiferromagnet on the triangular lattice, make it clear that a thorough theoretical study is required to begin understanding this material's complex behavior.  Here we attempt to access the global phase diagram of this system by analyzing the roles of three important known perturbations away from the `ideal' classical triangular antiferromagnet---quantum fluctuations stemming from the low spin $S=1/2$, spatial exchange anisotropy, and Dzyaloshinskii-Moriya (DM) interactions---all of which compete and favor different spin arrangements.  Because the ideal classical model exhibits a large ground state degeneracy, we show that even weak spatial anisotropy and DM coupling are sufficient to qualitatively alter the standard Y-UUD-V phase diagram, stabilizing new spin orders including incommensurate `umbrella' and planar states, a commensurate `distorted V' spin structure, and a commensurate `inverted Y' phase.  Further experiments (such as neutron scattering and NMR measurements) will be very helpful for identifying which of these orders appear in the observed phase diagram, thereby sharpening the outstanding theoretical questions that undoubtedly remain.  We also predict that---in sharp contrast to \ccc---the phase diagram of \ccb\ ought to be insensitive to the direction of the magnetic field inside of the triangular planes.

The remainder of the paper is organized as follows.  We provide an overview of the model and strategy that we pursue in Sec.\ \ref{ModelStrategy}, and review the physics of the isotropic triangular antiferromagnet in Sec.\ \ref{sec:isotropic-states}.  Section \ref{sec:DM} addresses the case where DM coupling is present but spatial exchange anisotropy is neglected.  We then explore the Bose-Einstein condensation of magnons near the saturation field in Sec.\ \ref{sec:bec}, which allows us to simultaneously treat quantum fluctuations, DM coupling, \emph{and} spatial anisotropy in a simple setting.  The influence of spatial anisotropy on the global phase diagram is studied with and without DM interactions in Secs.\ \ref{SpatiallyAnisotropicModel} and \ref{AnisotropicPlusDM}, respectively.  Finally, we provide a summary and concluding remarks in Sec.\ \ref{Conclusions}.

\section{Model and Strategy}
\label{ModelStrategy}

\begin{figure}
 \includegraphics[width=2.6in]{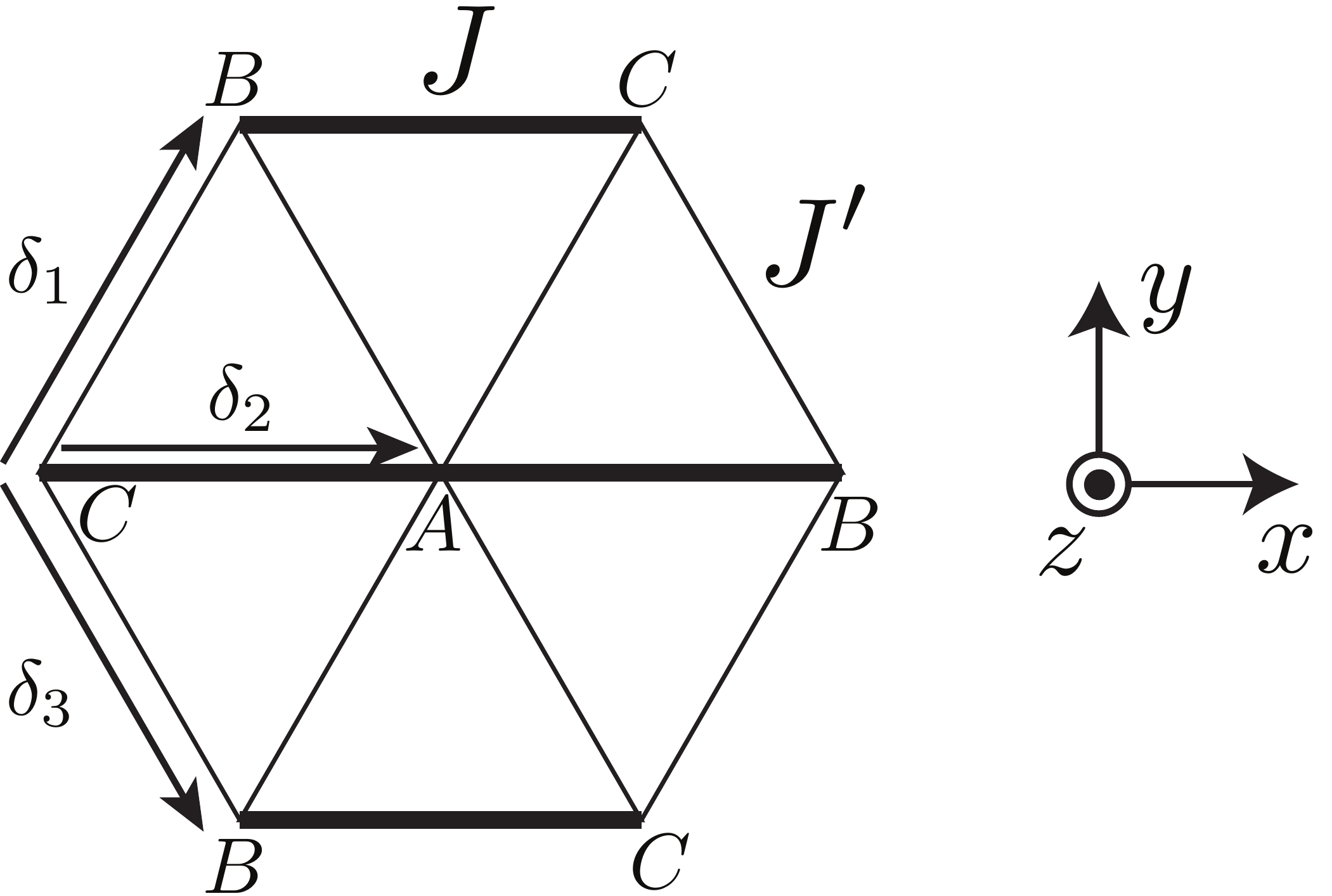}
  \caption{Spatially anisotropic triangular lattice with exchange $J$ along horizontal bonds and $J'$ along diagonal bonds.  We define axes such that the sites lie in the $(x,y)$ plane as shown.  Vectors ${\bm \delta}_{1,2,3}$ connect nearest-neighbor sites of the lattice.  In the isotropic limit where $J= J'$ and \DM~coupling is absent, the Hamiltonian in Eq.\ (\ref{Model}) exhibits a highly degenerate classical ground state manifold wherein spins order in an underconstrained three-sublattice pattern.  We label the three sublattices by $A,B$, and $C$.  } 
  \label{LatticeFig}
\end{figure}

We study the following Hamiltonian,
\begin{eqnarray}
   H &=& \sum_{\langle {\bf r r'} \rangle} J_{{\bf r,r'}} {\bf S}_{\bf r} \cdot {\bf S}_{\bf r'} - 
   \sum_r {\bf h} \cdot {\bf S}_{\bf r} + H_{\rm{DM}} 
\label{Model}
\end{eqnarray}
where the exchange integral $J_{\bf r,r'}$ is given by $J$ on the horizontal bonds and $J'$
on the diagonal zig-zag bonds as shown in Fig.\ \ref{LatticeFig}.  
For \ccb, experiments have measured the values\cite{tsujii2007} $J=11.3K$ and $J'=8.3K$. 
The second term describes the Zeeman energy
of spins in an external magnetic field while the third, $H_{\rm{DM}}$, represents the
asymmetric DM interaction between neighboring spins.  

Ideally, one would like to obtain the phase diagram for the quantum spin-1/2 problem above to begin understanding the interesting phenomenology of \ccb.  In this paper we will attempt to access the physics of the spin-1/2 system coming from the large-$S$ limit, including quantum fluctuations perturbatively.  This nevertheless still leaves a problem of substantial complexity, which can be understood by considering the classical, spatially isotropic limit of $H$, without DM coupling.  Let us denote this minimal classical Hamiltonian by $H_0$.  As noted in the introduction and reviewed in detail in Sec.\ \ref{sec:isotropic-states}, at finite magnetic fields $H_0$ exhibits a large `accidental' ground state degeneracy.  Quantum fluctuations (and thermal fluctuations at finite temperature) lift this ground state degeneracy, as do spatial anisotropy and DM coupling.  However, \emph{all of these ingredients compete with one another, favoring completely different ordered states}.  To resolve this competition, our strategy will be to compute the lowest-order energy splittings for the degenerate ground states of $H_0$ coming from each effect.  As we will see later quantum fluctuations and DM interactions split this degeneracy already at first order, while spatial anisotropy achieves this only at second order.  Thus despite the fact that in \ccb~spatial anisotropy [as quantified by $(J-J') \sim 0.3J$] is expected to greatly exceed the characteristic energy scales for DM interactions in the material, the two can in fact comparably influence the phase diagram.  

An additional complication arises from the fact that crystal symmetry of \ccb~permits several DM terms \cite{starykh10} which together break the SU(2) spin symmetry enjoyed by the exchange coupling down to a discrete subgroup.  
One might then expect the phase diagram to depend sensitively both on the polar angle that ${\bf h}$ 
makes with respect to the $z$-axis, normal to the triangular plane, \emph{and} the azimuthal angle 
${\bf h}$ makes in the $(x,y)$ plane.  Such a highly anisotropic phase diagram indeed emerges in the 
more spatially anisotropic material\cite{tokiwa2006} Cs$_2$CuCl$_4$.  In the limit of weak spatial 
anisotropy---which due to the relevance to \ccb~is our main focus here---most of these DM couplings fortunately play an unimportant role.  
As justified in Appendix \ref{sec:appDM}, it indeed suffices to consider only
\be
H_{\rm{DM}} =   -\sum_{{\bf r}} {\bf D} \cdot \left[{\bf S}_{\bf r} \times ({\bf S}_{{\bf r}+\bm{\delta}_1} + 
{\bf S}_{{\bf r} +\bm{\delta}_3})\right] ,
\label{eq:DM}
\ee
where the DM vector is ${\bf D} = D {\bf \hat{z}}$ (we assume $D>0$ throughout) and $\bm{\delta}_{1,3}$ are vectors shown in Fig.\ \ref{LatticeFig}.   
The strength of this DM interaction
is expected to be comparable to that in the isostructural material \ccc, where\cite{coldea2002}
$D\approx J/20$. 

Notice that at ${\bf h = 0}$ the model with the above DM coupling still exhibits a U(1) spin symmetry corresponding to global spin rotations about the ${\bf D}$ vector.  This leads to a major simplification---the phase diagram of $H$ depends on the polar angle ${\bf h}$ makes with the $z$-direction but \emph{not} on the azimuthal angle.  
All additional DM terms which lower this symmetry lift the accidental degeneracy of $H_0$ only at second order in their couplings, whereas $H_{\rm DM}$ above has a first-order effect; see Appendix \ref{sec:appDM} for details.  We emphasize that a nontrivial consistency check emerges here regarding relevance of our results to experiments on \ccb.  Our approach postulates that the physics of this material can be accessed coming from the classical, isotropic model without DM coupling.  If this postulate is correct, then contrary to Cs$_2$CuCl$_4$, the experimental phase diagram of \ccb~should be qualitatively insensitive to rotations of the field about the ${\bf \hat{z}}$ direction.  Recent experiments by Y.\ Takano and collaborators
\cite{takano_private} have indeed shown that the low-temperature phase diagram of \ccb~is qualitatively the same for fields oriented along the material's ${\bf b}$ and ${\bf c}$ axes ($x$ and $y$ axes in our notation from Fig.\ \ref{LatticeFig}).  This finding lends strong experimental support to our approach.

\section{Review of the Isotropic Model: $J = J', D = 0$} 
\label{sec:isotropic-states}

\subsection{Classical ground states}
\label{sec:iso_T0}

In the absence of DM coupling, the classical isotropic
Heisenberg model, where spins are described as 3D unit vectors, 
is well known to exhibit a large `accidental' ground
state degeneracy in applied magnetic fields \cite{kawamura1984}.
The classical ground state structure can be conveniently exposed by expressing the 
Hamiltonian (up to a constant) as
\begin{eqnarray}
   H_0 = \frac{J}{2}\sum_{\bf r} \left[{\bf M}_\Delta({\bf r}) -
   \frac{\bf h}{3J}\right]^2, \label{Hiso}
   \label{plaquethamiltonian}
\end{eqnarray}
where
\begin{equation}
   {\bf M}_\Delta({\bf r}) = {\bf S}_{\bf r} + {\bf S}_{{\bf r}+\bm{\delta}_1} +
   {\bf S}_ {{\bf r}+\bm{\delta}_2}
   \label{plaquetvariable}
\end{equation}
represents the magnetization of an elementary triangle located at site
${\bf r}$.  Below the saturation field $h_{\rm sat}^0 = 9J$, it is clear
from Eq.\ (\ref{Hiso}) that classical ground states satisfy the
condition ${\bf M}_\Delta({\bf r}) = {\bf h}/(3J)$ for all ${\bf r}$.
States fulfilling this requirement exhibit a three-sublattice
structure as shown in Fig.\ \ref{LatticeFig}, whose spins ${\bf S}_{A,B,C}$ are
determined from
\begin{equation}
  {\bf S}_A + {\bf S}_B + {\bf S}_C = \frac{{\bf h}}{3 J}.
  \label{GdStCondition}
\end{equation}

The ground state degeneracy follows immediately from 
Eq.\ \eqref{GdStCondition}, since the spins on the three sublattices are
specified by a total of six angles which are constrained by only three
equations. 
Parametrizing the spins on sublattice $\alpha$ as 
\begin{equation}
  {\bf S}_\alpha = (\sin\theta_\alpha\cos\phi_\alpha,\sin\theta_\alpha\sin\phi_\alpha,\cos\theta_\alpha),
  \label{Salpha}
\end{equation}
the angles must specifically satisfy
\begin{eqnarray}
 \frac{{\bf h}\cdot {\bf \hat{x}}}{3J}
 &=& \sin\theta_A\cos\phi_A + \sin\theta_B \cos\phi_B + \sin\theta_C \cos\phi_C  
  \nonumber \\
  \frac{{\bf h}\cdot {\bf \hat{y}}}{3J}
 &=& \sin\theta_A\sin\phi_A + \sin\theta_B \sin\phi_B + \sin\theta_C \sin\phi_C
  \nonumber  \\
  \frac{{\bf h}\cdot {\bf \hat{z}}}{3J}
 &=& \cos\theta_A + \cos\theta_B + \cos\theta_C 
\label{groundstateparameterization} 
\end{eqnarray}
Equations (\ref{groundstateparameterization}) allow one to, say, express $\phi_B, \phi_C$, and $\theta_C$ in terms of $\phi_A, \theta_A$, and $\theta_B$.  Note, however, that classical ground states do not exist for all possible values of the latter angles.

\begin{figure}
   \includegraphics[width=2.6in]{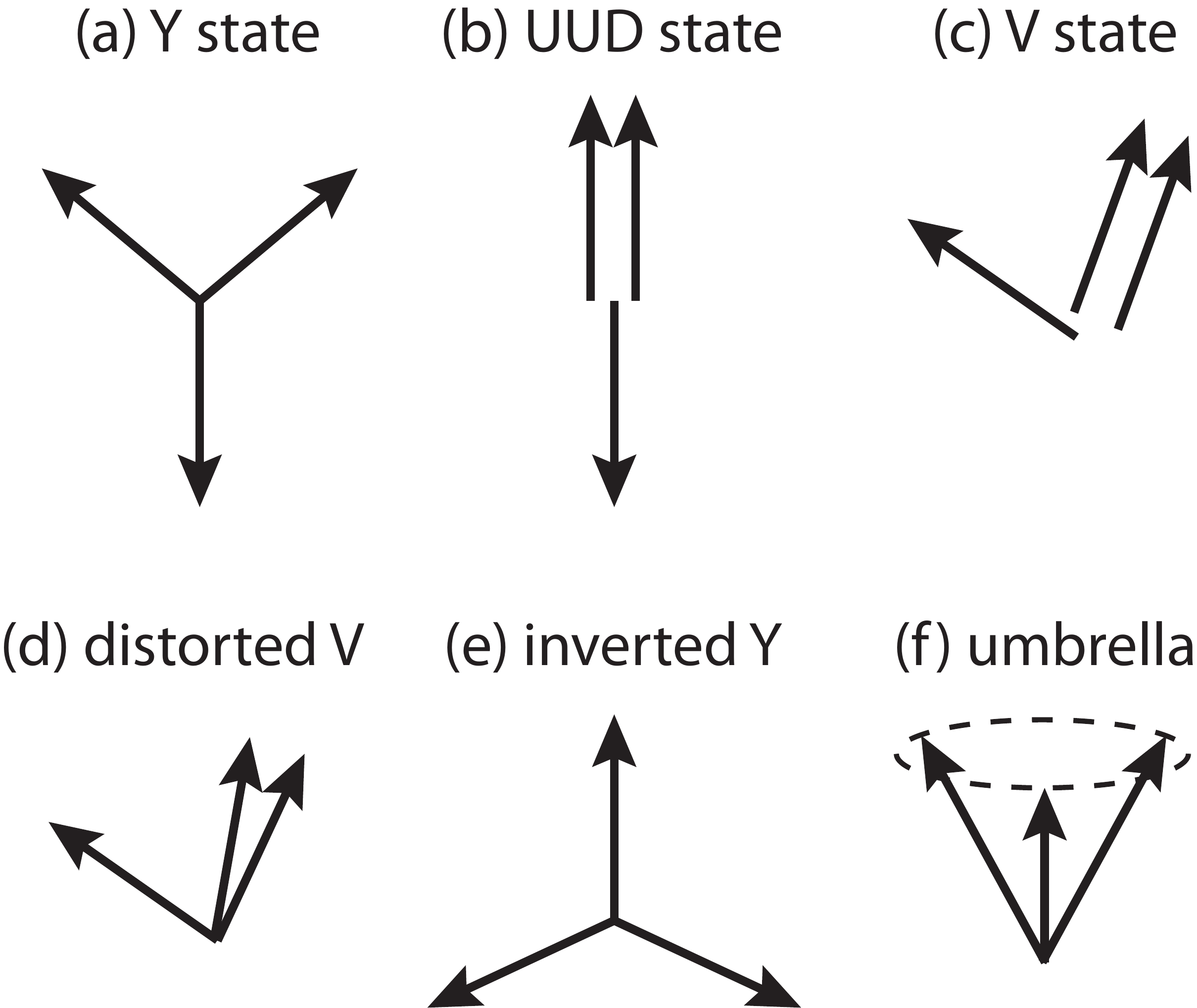}
   \caption{
     \label{GroundStatesfig}  Important members of the classical ground state manifold exhibited by Eq.\ (\ref{Model}) when $J = J'$ and \DM~coupling is absent: (a) Y, (b) up-up-down (UUD), (c) V, (d) `distorted V', (e) `inverted Y', and (f) umbrella states.  The arrows denote the spin orientations on the $A$, $B$, and $C$ sublattices of Fig.\ \ref{LatticeFig}.  }
     \label{ClassicalGroundStates}
\end{figure}

At ${\bf h} = 0$ where the classical Hamiltonian exhibits
O(3) symmetry, this degeneracy is symmetry-related: the three
unconstrained angles reflect the arbitrariness of the plane in which the $120^\circ$ spiral 
orients and the freedom for rotating all spins by an arbitrary angle within that plane.  Introducing
a finite magnetic field reduces the spin symmetry down to U(1).  The
three free angles nevertheless persist, only one of which is now
symmetry-related (corresponding to global rotations of the spins about
the magnetic field axis).  Consequently, in finite magnetic fields
below saturation, $H_0$ exhibits an accidental classical ground state
degeneracy specified by two continuously deformable angles.  
Figure \ref{GroundStatesfig} illustrates some important members of this
ground state manifold which we will frequently refer to later on: (a)
Y, (b) up-up-down (UUD), (c) V, (d) `distorted V', (e) `inverted Y', and (f) umbrella states.  

\subsection{Symmetries of the classical ground states}

\label{ClassicalGroundStateSymmetries}

It is important to emphasize the symmetry distinction between the above phases.  Because $\langle {\bf S}_{\bf r}\cdot {\bf h}\rangle$ is uniform in umbrella configurations, this order is exceptional among the classical ground states in that it preserves the original lattice translation symmetry [when followed by an appropriate global U(1) spin rotation].  All other classical ground states exhibit a nontrivial three-sublattice pattern of $\langle {\bf S}_{\bf r}\cdot {\bf h}\rangle$ and therefore spontaneously break discrete translation symmetry.  UUD order is also exceptional---it alone preserves global U(1) spin rotations about the field axis and therefore does not break any continuous symmetry.  Ultimately, the U(1) spin symmetry enjoyed by the UUD state permits the formation of a one-third magnetization plateau in this system.  In a boson representation of the spins, 
umbrella order corresponds to a superfluid phase, UUD order is a solid phase, and all other classical ground states constitute 
supersolids.\cite{heidarian2005,jiang2009,wang2009}

These `supersolids' can be further distinguished by symmetry.  Since in the V state spins on two sublattices, say A and B, are parallel, this order is symmetric under spatial $\pi/3$ rotations about a site on the C sublattice. This can also be phrased as invariance of the V state 
with respect to permutation of the A and B sublattices. The distorted V state, while smoothly connected to the V state, violates this symmetry 
(the A and B sublattices are not equivalent anymore)
and therefore constitutes a distinguishable phase in the isotropic system.  
Finally, in both the Y and inverted-Y states spins on one sublattice, say C, point either along or against the field, 
and both orders are invariant under $\pi/3$ rotations about a site on sublattice C followed by a global $\pi$ spin rotation.  
While the latter phases exhibit identical symmetries, one can not deform the spins smoothly from one configuration to the 
other without breaking additional symmetries; thus it is still meaningful to classify them as different states.  

All of the orders displayed in Fig.\ \ref{ClassicalGroundStates} thus constitute distinct phases when the full symmetries of the isotropic Heisenberg triangular antiferromagnet in field are present.  Note, however, that when the symmetry is lowered by including DM coupling and/or spatial anisotropy, this ceases to be the case.  The V and distorted V states, for example, are not distinguishable once $\pi/3$ rotation symmetry is removed by adding either of these ingredients.  Similarly, the distinction between the UUD and Y states hinges on U(1) spin symmetry and rotation symmetry, both of which are lost in the presence of DM coupling when the magnetic field has a finite component in the $(x,y)$ plane.  The UUD and distorted V states can, by contrast, remain distinct even in this physical situation, despite the very low symmetry remaining for the problem.  For example, when ${\bf h} = h{\bf \hat{x}}$ and DM coupling is non-zero the Hamiltonian remains invariant under $S^x_{\bf r} \rightarrow S^x_{-\bf r}$ and $S^{y,z}_{\bf r} \rightarrow -S^{y,z}_{-\bf r}$; the UUD state respects this discrete symmetry while the distorted V state does not.  We will return to these issues in subsequent sections of the paper.

\subsection{Influence of thermal fluctuations}
\label{sec:iso_finiteT}

\begin{figure}
\includegraphics[width=3.5in]{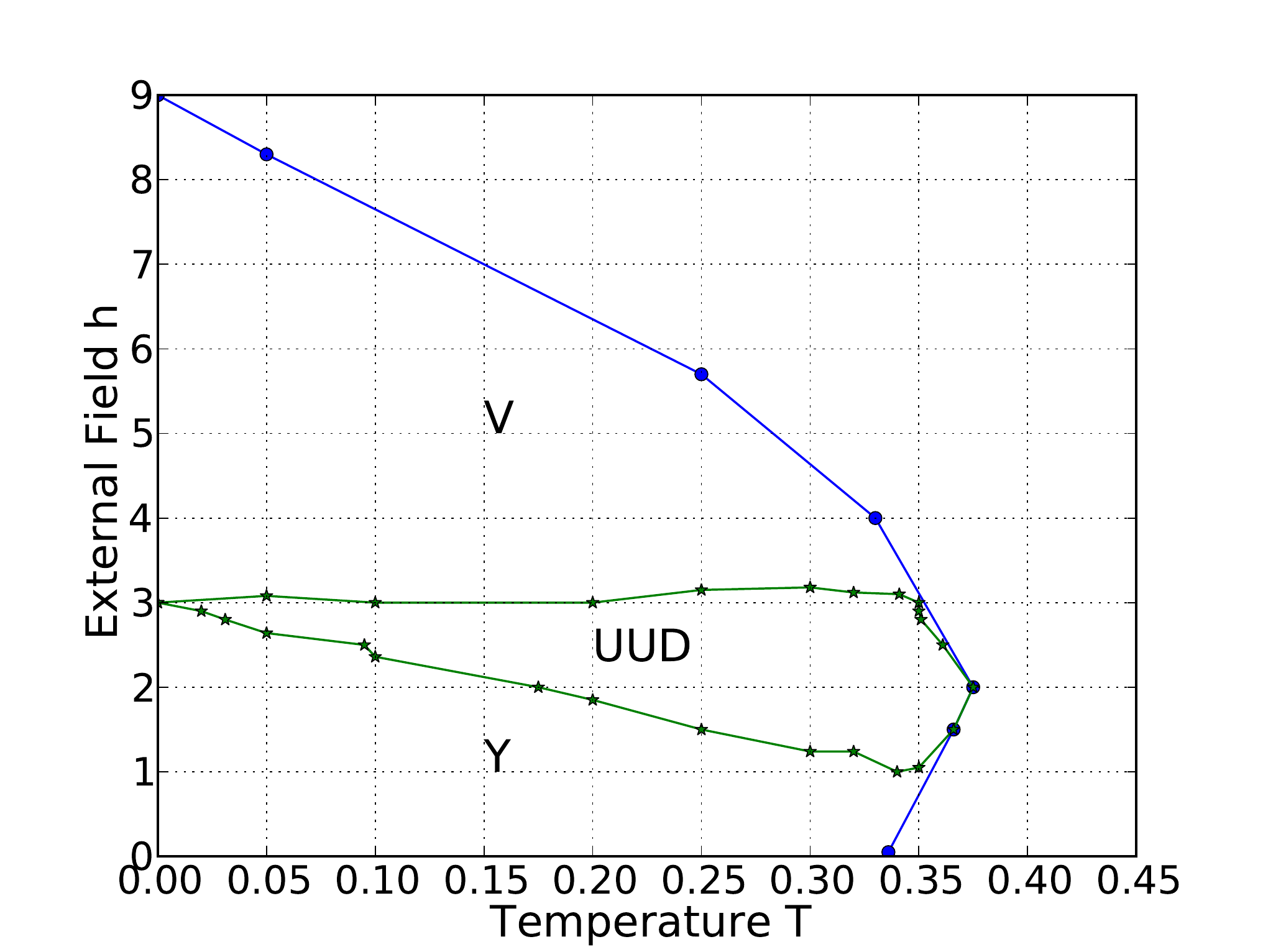}
\caption{(Color online) Phase diagram of the classical isotropic model.  Amongst the large set of accidentally degenerate classical ground states, thermal fluctuations select the Y, UUD, and V states illustrated in Figs.\ \ref{ClassicalGroundStates}(a) through (c).}  
\label{classisopd}
\end{figure}

Thermal fluctuations provide one mechanism that lifts the large accidental ground state degeneracy of the classical model.
At finite temperature the system minimizes its free energy, $F=U-TS$, and although the ground states exhibit identical energies $U$, their entropies generically differ.  Thus only the most entropically favored states emerge. The finite temperature phase diagram was established long ago\cite{kawamura1984}, and is reproduced with our numerics in Fig.\ \ref{classisopd}. 

This set of Monte-Carlo simulations is based on ALPS code.\cite{ALPS} 
System sizes ranged 
from $24\times24$ to $96\times96$ with most simulations being carried out on $48\times48$ lattices. 
(The triangular lattice can be viewed as a square lattice with bonds along one diagonal; the system sizes above refer to the dimensions of such a square lattice.  All simulations were performed with periodic boundary conditions.)
Every $(T,h)$ coordinate
was simulated using the standard Metropolis algorithm, with 
500,000 Monte Carlo steps for thermalization and another 500,000 for measurement in the $24\times24$ system.
For $48\times48$ lattices these numbers were in the range $150,000 - 250,000$.
All Monte Carlo data presented here and below correspond to $J  = 1$.  
Different phases and boundaries between them
have been identified by the behavior of the specific heat, the magnetization $M$ and its field derivative
$dM/dh$, and the vector chirality ${\bm \kappa}$ and coplanarity $K$ defined below.

The vector chirality is defined as
\bea
{\bm \kappa} &=& \frac{2}{3\sqrt{3}} \frac{1}{N} \sum_{\bf{r}} 
\Big({\bf S}_{\bf r} \times {\bf S}_{{\bf r} + \bm{\delta}_1} + 
\nonumber\\
&& + {\bf S}_{{\bf r} + \bm{\delta}_1} \times {\bf S}_{{\bf r} +\bm{\delta}_2} + 
{\bf S}_{{\bf r} + \bm{\delta}_2} \times {\bf S}_{\bf r}\Big) ,
\label{eq:chirality}
\eea
where $N$ is the total number
of spins in the system and the normalization factor ensures that the maximal chirality magnitude is unity.
Taking ${\bf h} = h{\bf\hat{z}}$ for concreteness in the remainder of this section, it is useful to consider the longitudinal $\kappa_z$ and
transverse $\kappa_\perp = (\kappa_x^2 + \kappa_y^2)^{1/2}$ components of the vector chirality, in addition to
its magnitude $\kappa = (\kappa_z^2 + \kappa_\perp^2)^{1/2}$.  Finite longitudinal chirality
$\kappa_z$ identifies non-coplanar spin structures of umbrella type while
a non-zero transverse chirality signifies coplanar ordering in which no two spins in the unit cell are parallel (such as Y states). Notice that by construction the chirality vanishes in three-sublattice
states containing two parallel spins, such as the UUD and V states.

The coplanarity $K$, introduced in Ref.\ \onlinecite{watarai01}, 
provides another useful indicator of planar spin structures. This quantity follows from the $A/B/C$ sublattice magnetizations
\be
{\bf M}_{A/B/C} = \frac{1}{N} \sum_{{\bf r} \in A/B/C} {\bf S}_{\bf r} 
\ee
by constructing
\be
{\bf K}_{AB} = ({\bf M}_A \times {\bf M}_B)\times{\bf\hat{h}}
\label{eq:coplanarity1}
\ee
and similarly for  ${\bf K}_{BC}$ and ${\bf K}_{CA}$.  The coplanarity is then given by the combination
\be
K^2 = |{\bf K}_{AB}|^2 + |{\bf K}_{BC}|^2 + |{\bf K}_{CA}|^2 .
\label{eq:coplanarity2}
\ee
As defined $K$ is finite for any coplanar spin configuration which includes the magnetic 
field in its plane. In particular, $K$ is finite in the V state but vanishes in the collinear UUD configuration.

\begin{figure}
\subfigure[]{\label{subfig:mag-a}\includegraphics[width=0.45\textwidth]{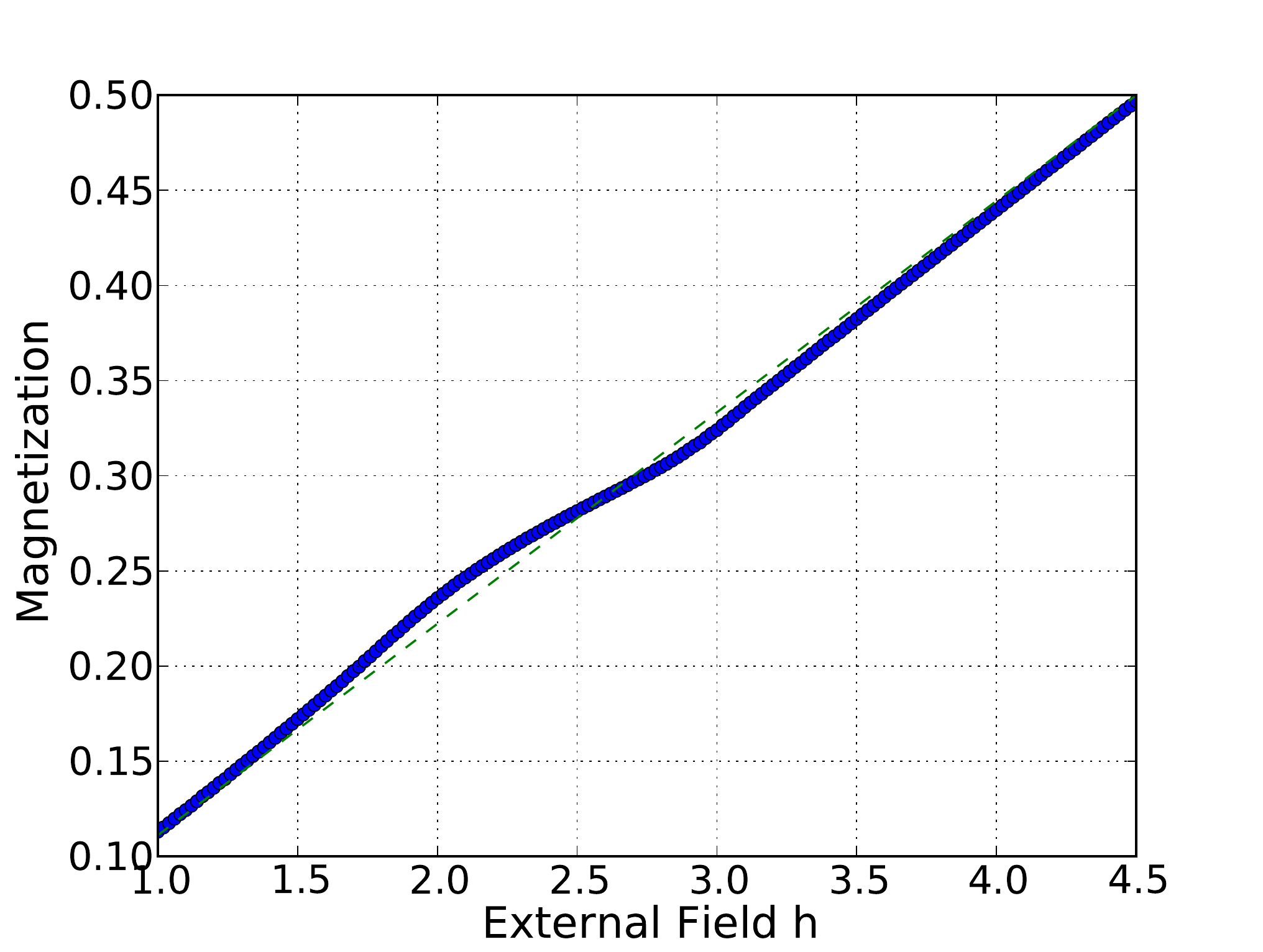}} 
\vspace{0.5cm}
\subfigure[]{\label{subfig:mag-b}\includegraphics[width=0.45\textwidth]{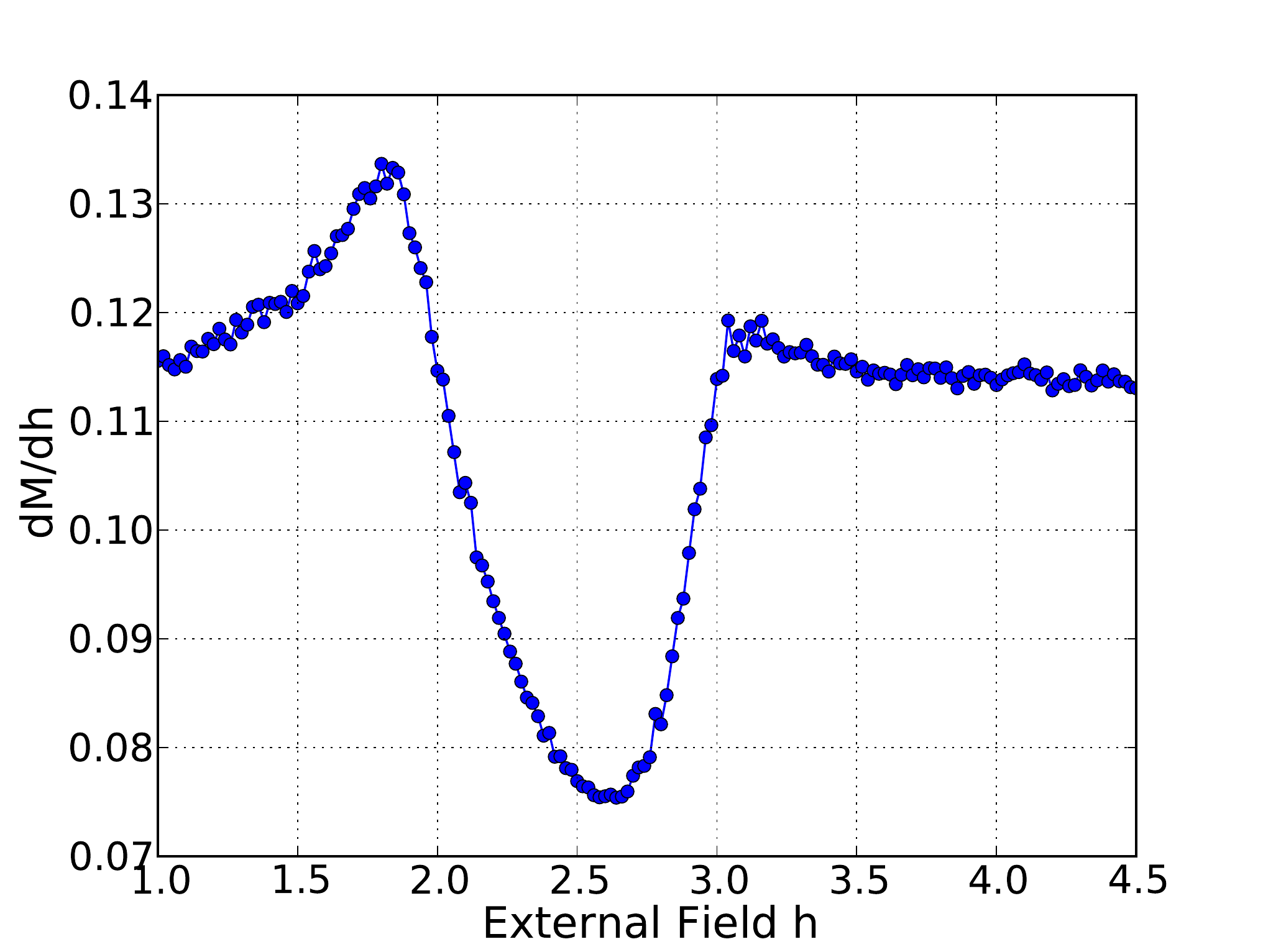}} 
\caption{(Color online) (a) Magnetization $M$ and (b) its field derivative $dM/dh$ in the vicinity of one-third magnetization.  The data were obtained with Monte Carlo simulations of the classical isotropic model at temperature $T = 0.2J$.  The collinear UUD state underlies the drop in magnetic susceptibility near $h = 2.5J$, visible in both (a) and (b).  The dashed line in (a) shows the $T= 0$ magnetization of the classical triangular antiferromagnet, $M=h/(9 J)$.}
\end{figure}

Our Monte-Carlo simulations, summarized in Fig.\ \ref{classisopd}, agree well with existing 
data \cite{kawamura1984,Gvozdikova2011} and clearly demonstrate
entropic selection of the Y, UUD, and V states over non-coplanar order.
This is consistent with the usual intuition that entropy disfavors non-coplanar structures and 
favors coplanar and, when available, collinear ordering\cite{Shender,Henley}.  
Coplanarity of the selected states is reflected in an exceedingly small value of the longitudinal chirality $\kappa_z$
over the entire magnetic field range $0 \leq h \leq 9J$.
Figures \ref{subfig:mag-a} and \ref{subfig:mag-b} exhibit the magnetization $M$ and its field derivative, $dM/dh$, at $T = 0.2J$ over the same field range.  
Notice that the UUD state is very well identified by the abrupt variation of $dM/dh$,
with two peaks in Fig.\ \ref{subfig:mag-b} representing the lower and upper critical fields. 
Away from the UUD state the slope of the magnetization approaches $1/(9J)$,
which is just the uniform susceptibility of the classical antiferromagnet at zero temperature.
The magnetization slope
in the UUD state [Fig.\ \ref{subfig:mag-a}] is visibly smaller, but
clearly remains finite.  This is a manifestation of the `dual' role played by temperature:
it simultaneously selects coplanar states \emph{and} thermally disorders them,
eventually resulting in a paramagnetic state above a field-dependent critical temperature.

Despite being such a well-studied problem, there remains a serious question about the
phase diagram of the isotropic triangular lattice antiferromagnet at ${\bf h} =0$. 
As described above in Sec.\ \ref{sec:iso_T0}, the large degeneracy of the model in this limit
is symmetry related and reflects arbitrariness of the ordering plane in which $120^\circ$ spiral forms.
This makes the order parameter space isomorphic to SO(3), the group of rotations
of a three-dimensional rigid body \cite{kawamura1984}.
According to the Mermin-Wagner theorem, this continuous symmetry cannot be broken
at any finite temperature. Nonetheless Monte-Carlo simulations, including ours, do show a
weak peak in specific heat at finite temperature, approximately $0.33 J$. It has been suggested \cite{kawamura1984}
that this finite-T feature in fact reflects a non-trivial phase transition associated 
with binding of Z$_2$ vortices permitted by the SO(3) structure.
At present our study, which aims to understand the global features of the deformed Heisenberg model at
finite $T$ and ${\bf h}$, has nothing to add to this interesting issue \cite{wintel1995,kawamura2007,mzh2009,kawamura2010A,kawamura2010B,Gvozdikova2011}, 
the resolution of which requires more extensive Monte-Carlo simulations.

We should emphasize, however, that at {\em finite} magnetic fields the entropically selected states break 
translational symmetry and introduce three inequivalent sublattices A, B and C.    
Consequently, at any finite ${\bf h}$ up to saturation there exists a finite critical temperature $T_{\rm d}$ 
arising from discrete symmetry breaking associated with the spontaneous selection of the sublattice ordering.  
In Fig.\ \ref{classisopd} this hull-shaped $T_{\rm d}(h)$ separates low-temperature ordered states
from a high-temperature paramagnetic phase in which only the uniform magnetization $M$ is present.

The finite-field situation contains yet another unresolved issue, related to the U(1) symmetry corresponding 
to rotations about the magnetic field direction.  The coplanar
Y and V states break this 
symmetry by selecting an ordering direction in the plane perpendicular to
the field direction.  While due to dimensionality the U(1) symmetry can not be truly broken at finite temperature, 
there exists a Kosterlitz-Thouless (KT) temperature $T_{\rm KT}$ below which the transverse magnetic order 
exhibits power-law correlations.  In principle one could have $T_{\rm KT} \neq T_{\rm d}$ (see Ref.\ \onlinecite{mzh2009}
for a recent example of such behavior).  From the available data, however, 
it appears that within our numerical accuracy the two transitions coincide, $T_{\rm KT} = T_{\rm d}$.  
A related issue is that at finite temperature, the transitions from UUD to either Y or V are also of KT type, and as such are inherently broad and difficult to locate precisely in numerics \cite{transitions}. 
We would like to emphasize again that our main focus is on identifying possible phases rather than 
studying the nature of the transitions between them.  
Readers interested in properties of such phase transitions in a similar 
context are instead referred to the recent detailed study in Ref.\ \onlinecite{transitions}.

\subsection{Quantum ground states}

\label{QuantumEffectsIsotropicLimit}

As first shown by Chubukov and Golosov\cite{ChubukovGolosov}, the accidental degeneracy inherent in the classical isotropic model can be lifted even at zero temperature by quantum fluctuations.  Here zero-point energy of spin waves---rather than entropy---provides the degeneracy-lifting mechanism.  Quantum ground state selection can be systematically explored using the machinery of the $1/S$ expansion.  To this end, one starts with a particular classical ground state in the degenerate manifold and defines rotated spin coordinates $\tilde{\bf S}_{A/B/C}({\bf r})$ on the three sublattices such that the classical order corresponds to $\langle \tilde{\bf S}_{A/B/C}({\bf r})\rangle = S{\bf \hat{z}}$.  One then introduces Holstein-Primakoff bosons $a({\bf r}),b({\bf r})$, and $c({\bf r})$ to express the rotated spin operators to leading order in $1/S$ as
\begin{eqnarray}
  \tilde S_A^x({\bf r}) &=& \sqrt{\frac{S}{2}}[a({\bf r}) + a^\dagger({\bf r})]
  \nonumber \\
  \tilde S_A^y({\bf r}) &=& i\sqrt{\frac{S}{2}}[a({\bf r}) - a^\dagger({\bf r})]
  \nonumber \\
  \tilde S_A^z({\bf r}) &=& S-a^\dagger({\bf r})a({\bf r})
  \label{eq:HP}
\end{eqnarray}
and similarly for $\tilde {\bf S}_{B/C}({\bf r})$.  Retaining only the leading-order terms in $1/S$ generates a Hamiltonian of the form $H = E_0 + H_2$, where $E_0$ is the classical ground state energy and $H_2$ is quadratic in the three boson fields.  

\begin{figure}
\includegraphics[width=3in]{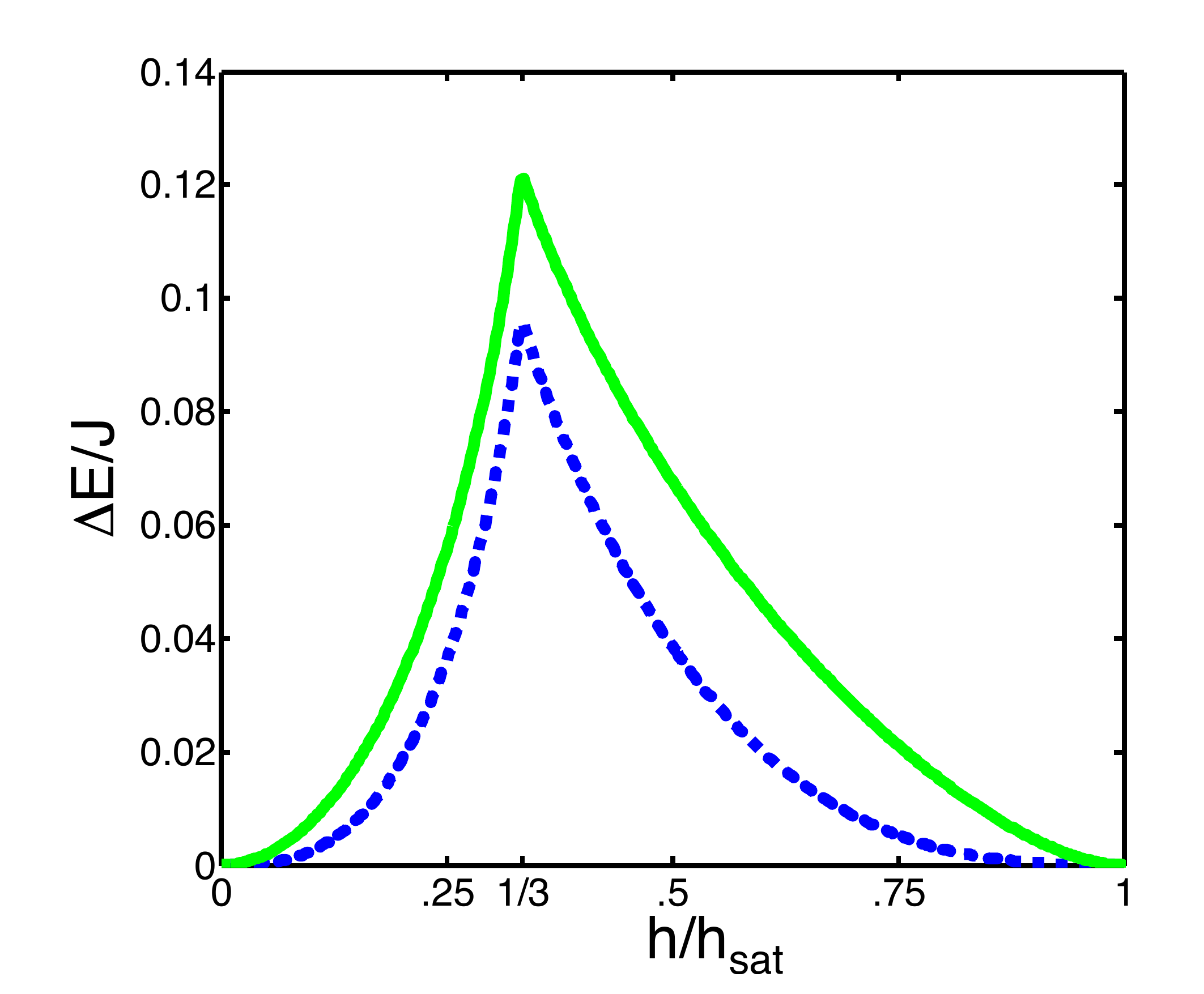}
\caption{(Color online) Leading $1/S$ quantum energy splittings between the Y/UUD/V states 
and the inverted-Y (dashed line) and umbrella states (solid line).  These curves illustrate an important trend --- 
quantum fluctuations are most effective at splitting the accidental degeneracy 
near 1/3 magnetization where low-energy collinear states are available.}
\label{quantenergydiff}
\end{figure}

Although at a given magnetic field $h$, $E_0$ is identical for all allowed classical ground states 
satisfying Eqs.\ \eqref{groundstateparameterization}, the spin-wave Hamiltonian $H_2$ \emph{does} depend on 
the particular ordered configuration around which one is expanding, via the dependence of the spin-wave
dispersion $\omega_k$ on the angles $\theta_{A/B/C}$ and $\phi_{A/B/C}$.  The zero-point energy for spin-waves,
$\langle H_2 \rangle = (1/2)\sum_k \omega_k$, therefore yields a $1/S$ correction to the energy that lifts the classical degeneracy. 
See, for example, Ref.\ \onlinecite{nikuni2} for an earlier application of this strategy to a related problem. 
We note that this analysis can be carried out rather efficiently by parametrizing the original spin operators 
via Eq.\ (\ref{Salpha}) and deriving the general form of $H_2$ as outlined above.  
The constraints in Eqs.\ (\ref{groundstateparameterization}) allow one to express $\phi_B$, $\phi_C$, and $\theta_C$ in terms of 
$\phi_A$, $\theta_A$, and $\theta_B$; due to the U(1) spin symmetry that 
is present when ${\bf h} \neq 0$, one can arbitrarily set $\phi_A = 0$.  
Consequently, $H_2$ can be expressed in terms of only two angles, 
$\theta_{A/B}$.  Diagonalizing $H_2$ for a mesh of $\theta_A$ and $\theta_B$ then allows one to compute the zero-point energies for the classical ground state manifold to deduce which states quantum fluctuations favor.  Although the selection for the isotropic model is already well-known, we will use this scheme later on to explore the competition between quantum fluctuations and DM interactions.  

It is important to emphasize a subtle point in the procedure outlined above.  At order $1/S$ quantum fluctuations not only split the degeneracy through zero-point motion, but also renormalize the spin structure in the classical ground states.  Thus to order $1/S$ the spin angles 
should read $\theta_A = \theta_A^0 + \delta\theta_A$, and similarly for other angles, 
where $\theta_A^0$ corresponds to a classical ground state and $\delta\theta_A$ is a $1/S$ quantum correction 
(which we do not calculate here).  When evaluating the zero-point energies $\langle H_2\rangle$ to order $S$, 
clearly one can safely neglect such quantum corrections to the spin states.  One may worry, however, that since the classical ground state energy $E_0$ is proportional to $S^2$, evaluating $E_0$ with the renormalized spin angles may produce an order-$S$ contribution to the energy that is comparable to the leading zero-point energy.  This is certainly not the case---because the classical ground states minimize $E_0$, 
the shifts $\delta \theta_A$ appear only at \emph{second order} and therefore contribute to the energy only at order $S^0$; see, \emph{e.g.}, the discussion in Ref.\ \onlinecite{zhitomirsky98}.  
Thus while renormalization of the spin structure is crucial for capturing certain physical quantities such as quantum corrections to the 
magnetization curve as a function of field, this effect is indeed unimportant for our main purpose: 
evaluating the ground state energies to leading order in $1/S$.

As in the case of thermal fluctuations, quantum fluctuations disfavor non-coplanarity and prefer collinear order, leading to Y states below 1/3 magnetization ($0 < h < h_{\rm sat}/3$), UUD ordering at 1/3 magnetization ($h = h_{\rm sat}/3$), 
and V states at larger magnetization ($h_{\rm sat}/3 < h < h_{\rm sat}$) up to saturation\cite{ChubukovGolosov}.  
While it is well known that the UUD phase realizes a magnetization plateau also in the quantum problem\cite{ChubukovGolosov}, we emphasize that this fact arises from higher-order quantum corrections such as the $S^0$ effect discussed above.
One can recover this feature by incorporating $1/S$ quantum renormalization of the equilibrium angles for the Y and V states, which broadens the UUD state into a plateau \cite{ChubukovGolosov}.  This result can be also be seen in the following complementary way.  Since the UUD state is collinear it does not break any continuous symmetry, and therefore exhibits only gapped spin excitations when one renormalizes the spin-wave spectrum to leading order in $1/S$ (note that such a renormalization shifts the ground state energy only at order $S^0$).  Thus a finite change in the magnetic field is required to destabilize this phase.  

To establish intuition that will prove beneficial in subsequent sections, it is useful to now ask how robust the quantum selection of Y/UUD/V order is at different magnetic fields.  One can quantify this by computing the first $1/S$ quantum correction to the energy, \emph{i.e}., evaluating $\langle H_2\rangle$, for the Y/UUD/V states and comparing it with the corresponding correction 
for other classical ground states which quantum fluctuations disfavor.  The size of these energy differences 
$\Delta E = (1/2) \sum_k [\omega_k^{\rm other} - \omega_k^{\rm Y/UUD/V}]$ provide one with a rough guide for how robust the quantum ground state selection is to the inclusion of other perturbations, such as DM interactions.  
Figure \ref{quantenergydiff} presents the energy splittings $\Delta E$ obtained for the umbrella 
(solid line) and inverted-Y states (dashed line) as a function of $h$. 
To produce the data shown we have extrapolated large-$S$ corrections to $S = 1/2$.
It is seen that Y/UUD/V order is preferred over umbrella and inverted-Y states for all values of the field.
It is also seen that being co-planar, the inverted-Y state has lower energy than the umbrella one.
Although only two states, umbrella and inverted-Y, are considered, the figure clearly illustrates a general trend which 
recurs throughout the paper---\emph{quantum fluctuations are most effective at selecting ground states in the 
vicinity of 1/3 magnetization where low-energy collinear spin configurations are available}. 

\subsection{Biquadratic approximation of quantum fluctuations (zero-point motion)}
\label{sec:biquadraticapproximation}

\begin{figure}
\includegraphics[width=3in]{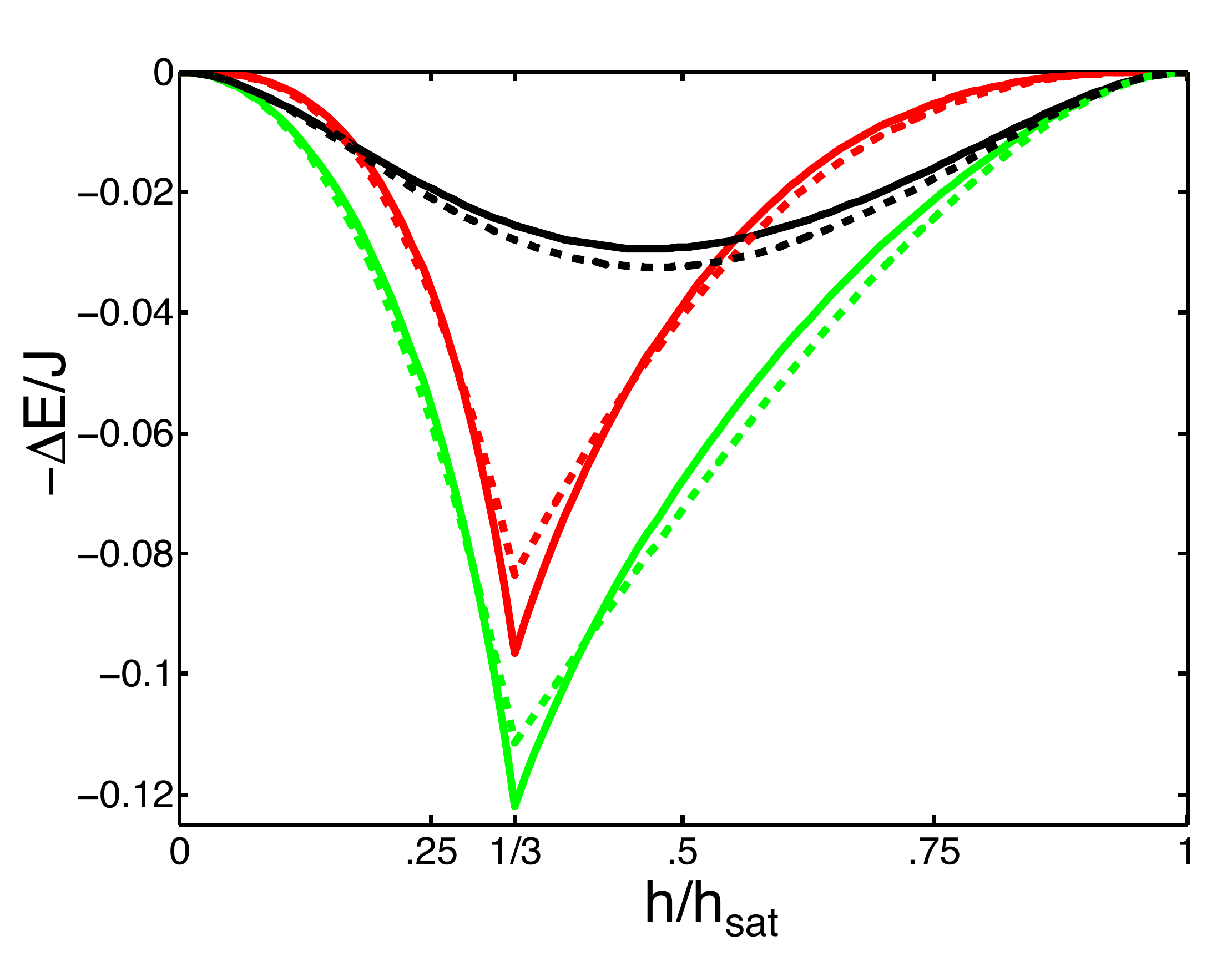}
	\caption{(Color online) Energy differences between various states using the leading $1/S$ result from 
	spin-wave theory (solid lines) and the biquadratic classical approximation (dashed lines).  
	Green lines represent the difference between Y/UUD/V and inverted Y energies; 
	red between Y/UUD/V and umbrella; and black between inverted Y and umbrella.}
	\label{Energydifferences}
\end{figure}

There is an alternative, albeit less rigorous, way to account for the influence of quantum fluctuations, within a purely \emph{classical} spin Hamiltonian.  This approach is based on the observation that coplanar and especially collinear spin states---which quantum fluctuations tend to favor most strongly---can be stabilized classically upon adding a suitable biquadratic spin coupling to the Hamiltonian\cite{ChubukovGolosov}.
This biquadratic interaction represents an effective Hamiltonian which is generated by quantum and/or thermal fluctuations \cite{henley2001}.
We have incorporated such a biquadratic spin interaction 
and optimized its (field dependent) coupling constant such that we 
reproduce the leading $1/S$ energy differences between several important states in the classical ground state manifold.   
In this way we find that quantum zero-point energy can be semi-quantitatively described by the following classical biquadratic
Hamiltonian with negative coupling constant,
\begin{eqnarray}
 \label{BiQuad}
  \delta H_{\rm bi} &=& - \Delta(h)\sum_{\langle{\bf r r'}\rangle} \left({\bf S}_{\bf r} \cdot {\bf S}_{\bf r'}\right)^2 , \\
    \Delta(h) &=& {\frac{0.0268}{S}}\left(1-0.03h\sqrt{h_{\rm sat}-h}\right).\nonumber
\end{eqnarray} 
Note that spins ${\bf S}_{\bf r}$ appearing in $\delta H_{\rm bi}$ are classical unit vectors
and that all dependence on the magnitude of the microscopic spin $S$ is contained in the field-dependent
coupling constant $ \Delta(h)$.
Figure \ref{Energydifferences} compares the energy differences between various classical ground 
states specified in the caption using the leading $1/S$ results (solid lines) and the biquadratic classical approximation 
(dashed lines), showing the quite good quantitative agreement between the two approaches.  

The full benefit of approximating quantum zero-point energy by the classical biquadratic
interaction will become clear when we address the effects of spatial anisotropy
$J' \neq J$ in Secs.\ \ref{AnisotropicQuantum} and \ref{sec:AnisotropicPlusDMquantum}. The classical energy gain
due to spatial anisotropy is quadratic in the difference $J-J'$.
Such second-order corrections are not in general easy to account for analytically, making it difficult 
to consistently compare the anisotropy-induced energy splittings with those arising from quantum effects and DM interactions.
Using the biquadratic interaction in Eq.\ \eqref{BiQuad}, however, allows us to circumvent this problem and
find the result of the competition between the different perturbations by numerically studying,
via standard Monte Carlo techniques, the classical system described by 
$H_0 + H_{\rm DM} + \delta H_{\rm bi}$.  Later we will refer to ground states obtained within such a classical model with biquadratic spin couplings as `pseudo-quantum ground states' to emphasize that despite corresponding to classical spins, such states are expected to semi-quantitatively reflect the influence of quantum fluctuations.  

\begin{figure}
\includegraphics[width=3in]{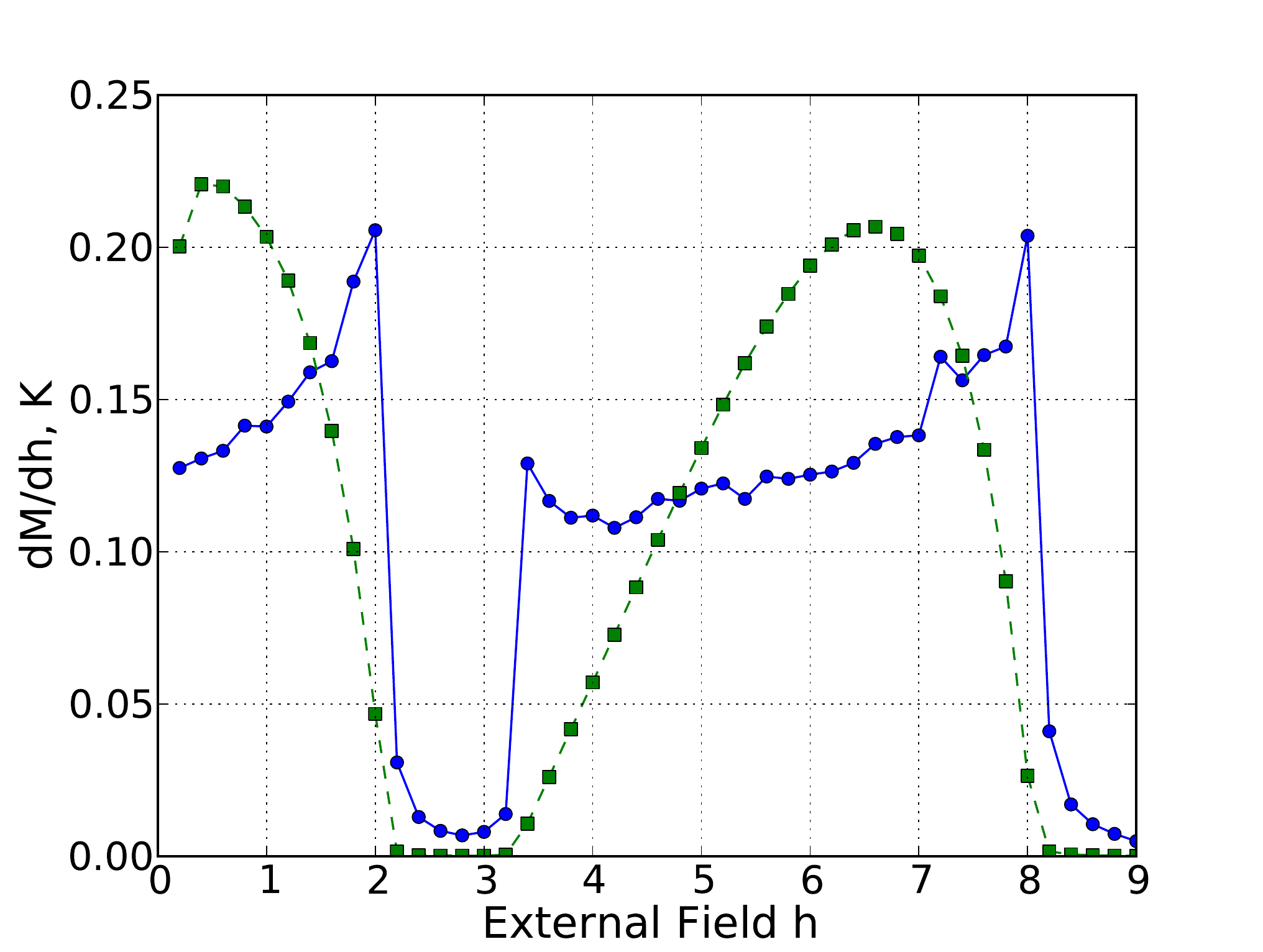}
	\caption{(Color online) $dM/dh$ (blue circles) and coplanarity $K$ (green squares), defined in Eq.\ \eqref{eq:coplanarity2}, 
	for the classical ``isotropic plus biquadratic'' model at $T=0.027J$,
	obtained by annealing from high temperature.}
	\label{DMDH_iso_bi}
\end{figure}

As a warm-up exercise of such a strategy, we first simulate a simpler ``isotropic plus biquadratic" Hamiltonian 
$H_{\rm iso+bi} = H_0 + \delta H_{\rm bi}$. Figure~\ref{DMDH_iso_bi} shows $dM/dh$ and the coplanarity $K$ for this model at a low temperature of $T=0.027J$.
This low temperature was reached via 
simulated annealing to avoid becoming trapped in local (but not global) free-energy minima.  
Specifically, Monte Carlo simulations were performed on a $24\times 24$ system starting at $T= 0.4 J$, 
then cooling down in small $\Delta T = 0.009J$ steps.  
At each temperature the system was equilibrated with $35,000$ Monte Carlo steps.  
Measurements were taken during the last 30,000 steps.

The pronounced minimum of $dM/dh$ visible in Fig.\ \ref{DMDH_iso_bi} over the field interval $2.2J \lesssim h \lesssim 3.2J$ identifies this
region with the collinear UUD state.  The coplanar nature of the adjacent phases
is evident from the plot of $K$ vs.\ $h$ in the same Figure. 
Note that despite the very low temperature considered, a substantial UUD plateau remains whose width 
greatly exceeds that of the purely classical model without interactions; see Fig.\ \ref{classisopd} for comparison.  
The width of the plateau is in fact proportional to $\Delta(h)$ as discussed in Ref.\ \onlinecite{ChubukovGolosov}:
quantum fluctuations, modeled here by $\delta H_{\rm bi}$, produce a stable UUD magnetization plateau over a finite field interval, even at $T = 0$.  
We note that the usual spin-wave approach predicts a plateau over a field interval of width $\Delta h \approx 1.8J/(2S)$, which is somewhat larger than the plateau width obtained in our isotropic-plus-biquadratic simulations.

\section{Spatially Isotropic Model with \DM~Interactions: $J = J', D \neq 0$}
\label{sec:DM}

In this section we will explore how DM coupling affects the classical and quantum phase diagrams discussed above.  One complication that arises here is that once we invoke DM interactions, the magnetic field direction is no longer arbitrary as it was in the previous section.  Recall that the DM vector points along the $z$-axis.  We will begin in Sec.\ \ref{sec:DMz} with the case where the field is applied along the $z$-direction, perpendicular to the triangular lattice plane.  This field orientation is simplest to analyze since the Hamiltonian including $H_{DM}$ then preserves global U(1) spin symmetry about the $z$-axis.  In-plane field orientations are studied in Sec.\ \ref{sec:DMx}.  

We should also emphasize that in any field orientation, DM coupling breaks $\pi/3$ rotation symmetry since $H_{\rm DM}$ in Eq.\ (\ref{eq:DM}) only couples spins on neighboring `chains' of the lattice.  Thus the model with only DM coupling (and no spatial anisotropy) is clearly fine-tuned.  Nevertheless, by examining the effects of DM alone we will gain valuable intuition that will be useful when we include the additional complication of spatial anisotropy in Sec. \ref{AnisotropicPlusDM}.

\subsection{Parallel orientation: ${\bf D} \parallel {\bf h} = h {\bf \hat{z}}$}
\label{sec:DMz}

\subsubsection{Classical Ground States}

We will first explore how DM interactions lift the accidental degeneracy of the classical isotropic model using first-order perturbation theory.  At this order, one simply needs to evaluate the DM energy using the unperturbed form of the classical ground states found in the previous section.  Using the parametrization for the ground states in Eq.\ (\ref{Salpha}) and the constraints in Eqs.\ (\ref{groundstateparameterization}), we obtain
\begin{eqnarray}
  \langle H_{\rm DM} \rangle &=& 2\sigma ND \bigg{\{}-\sin^2\theta_A\sin^2\theta_B 
  \nonumber \\
  &+& \frac{1}{324J^4}[h^2 + 9J^2 
  -6J h (\cos\theta_A + \cos\theta_B) 
  \nonumber \\
  &+& 18J^2 \cos\theta_A\cos\theta_B]^2\bigg{\}}^{1/2},
  \label{DMenergy}
\end{eqnarray}
where $\sigma = {\rm sign}[\sin(\phi_A-\phi_B)]$ determines the chirality of the spins in the $(x,y)$ plane.  The DM energy above is minimized when $\cos\theta_A = \cos\theta_B = \frac{h}{9J}$ and $\sigma = -1$.  One can readily see using Eqs.\ (\ref{groundstateparameterization}) that this corresponds to umbrella states of Fig.\ \ref{ClassicalGroundStates}(f) for all fields (with a specific chirality because $D \neq 0$ breaks inversion symmetry).  This outcome is quite natural given that DM coupling at ${\bf h} = 0$ favors coplanar spiral order with spins pointing in the triangular lattice plane; the field simply cants the spiral out of the plane, producing an umbrella pattern.  Note that this is markedly different from the coplanar and collinear configurations favored by thermal and quantum fluctuations.  This leads to an interesting competition once one incorporates these additional ingredients, as we now address.

\subsubsection{Influence of thermal fluctuations}
\label{ThermalFluctuationsDM}

One can deduce the qualitative outcome of the competition between DM interactions and thermal fluctuations at finite temperature using the following general argument.  In the previous section we discussed that while at $T = 0$ the isotropic model exhibits many classically degenerate ground states, at finite temperature entropy lifts this degeneracy in favor of Y/UUD/V states.  Moreover, this selection is strongest at intermediate temperatures where thermal fluctuations provide a large free-energy splitting but are not so severe as to destroy ordering altogether.  (This can be seen most clearly through the width of the UUD plateau in Fig.\ \ref{classisopd}, which reaches at maximum at around $T = 0.35J$ before entering the paramagnetic phase at slightly higher $T$.)  

Now suppose one slowly turns on DM interactions.  At zero temperature any finite DM coupling will immediately select umbrella ordering since the entropic splittings vanish as $T \rightarrow 0$.  By continuity umbrella states must persist over a finite temperature window that depends on both field and the DM coupling strength. Similarly, at intermediate temperatures the entropically favored Y/UUD/V states must also by continuity survive some amount of DM coupling.  This leads to multi-stage ordering as a function of temperature.  Consider for illustration the interesting example of one-third magnetization with weak DM coupling.  As temperature increases from zero the system will begin in an umbrella state, transition into an UUD plateau at intermediate temperatures, and eventually give way to a paramagnet at high temperatures.  

The qualitative picture that emerges is that as one increases DM coupling from zero, DM interactions eat away at the entropically-stabilized regions from the low-temperature side, eventually removing the Y/UUD/V phases entirely in favor of umbrella order at sufficiently large coupling.  Obtaining a more quantitative understanding of this competition requires extensive Monte Carlo simulations, which we will not explore here.  We will instead perform such a study in the more physically relevant case with spatial anisotropy in Sec.\ \ref{sec:anisotropy-classical}, where very similar physics arises.  It should also be kept in mind that in this subsection we neglected quantum effects entirely.  Below we incorporate quantum fluctuations and show that they lead to a substantially richer zero-temperature phase diagram than that of the purely classical model.  

\begin{figure}
\includegraphics[width=3.5in]{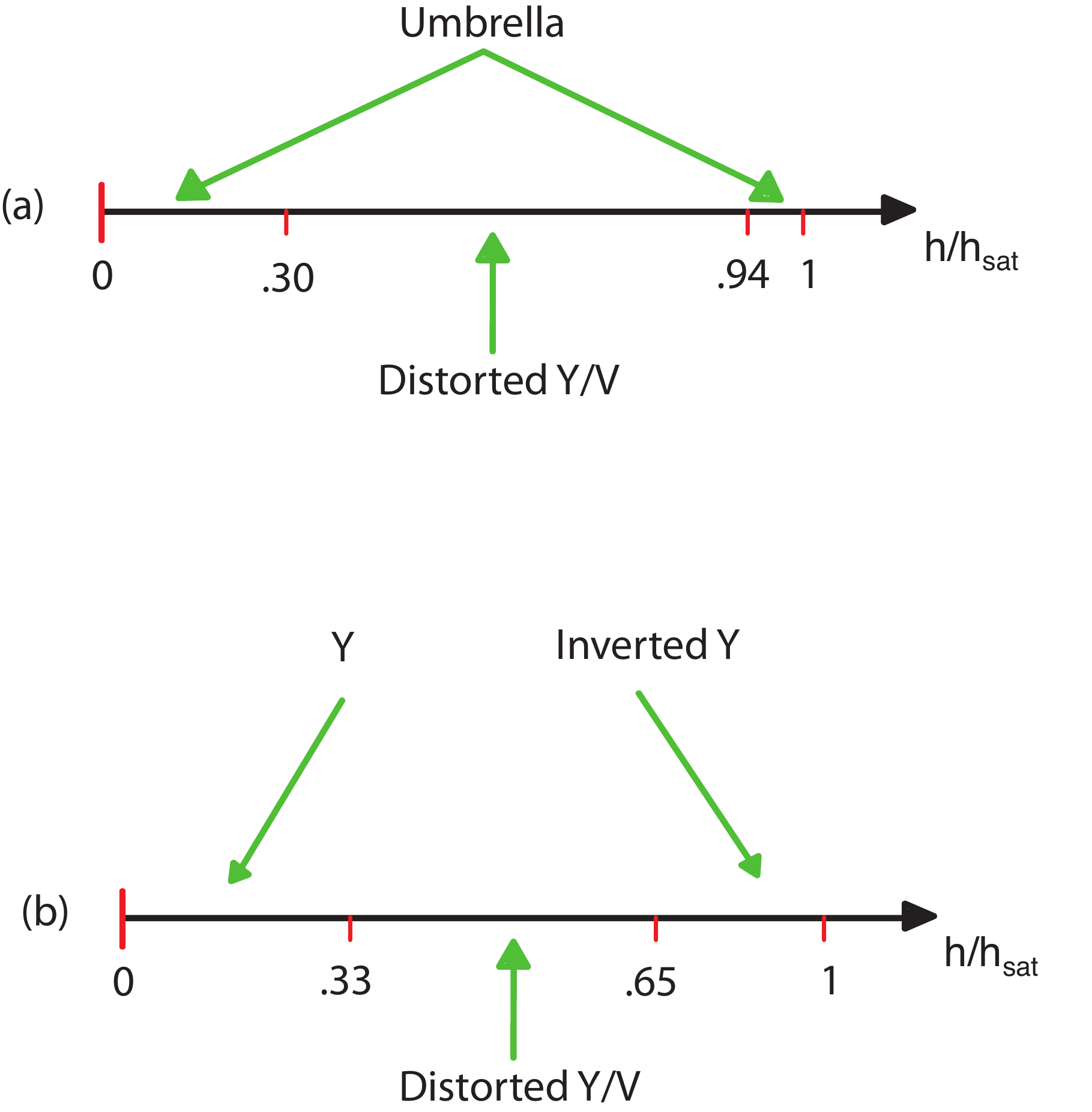}
	\caption{(Color online) Phase diagram of the quantum Heisenberg model with DM interactions, as a function of magnetic field $h$.
	The magnetic field orients along the DM axis, ${\bf D} \parallel {\bf h}$, in (a) and orthogonal to the DM axis,
	${\bf D} \perp {\bf h}$, in (b).}
	\label{fig:quantum+dm}
\end{figure}

\subsubsection{Quantum ground states}
\label{sec:DMz_quantum}

To incorporate DM interactions and quantum fluctuations at $T = 0$, we treat $1/S$ and $D/J$ as expansion parameters of the same order of magnitude.  The first-order corrections to the classical ground state energies can then be obtained simply by adding the $1/S$ spin-wave contribution discussed in Sec.\ \ref{QuantumEffectsIsotropicLimit} to the classical DM energy in Eq.\ (\ref{DMenergy}).  [For example, the effect of quantum fluctuations on the DM energy produces a higher-order correction $\sim(D/J)(1/S)$.]  By evaluating the perturbed energies for a dense subset of classical ground states as described in Sec.\ \ref{QuantumEffectsIsotropicLimit}, one can determine the spin order selected when both competing effects are present.  

Carrying out this procedure with $D/J = 0.05$ and $S = 1/2$, we have found that the umbrella phase is the quantum ground state 
for sufficiently small magnetization
($h/h_{\rm sat} \lesssim 0.3$) and, also, sufficiently near saturation ($h/h_{\rm sat} \gtrsim 0.94$); 
see  Fig.\ \ref{fig:quantum+dm}(a)  which summarizes 
our numerics for this case. In between, however, a different ground state 
emerges which reflects a compromise between DM interactions and quantum 
fluctuations---the distorted V phase of Fig.\ \ref{ClassicalGroundStates}(d). This state resembles the Y and V orders favored by quantum fluctuations, 
but is distorted in a non-coplanar fashion to gain DM energy.  Note that the distortion of the spins away from perfect Y or V order is not unexpected: these states, as well as UUD, exhibit rotation symmetry which is explicitly broken by the DM interaction in Eq. (\ref{eq:DM}) (recall the discussion at the end of Sec.\ \ref{ClassicalGroundStateSymmetries}).  
The analytical structure of the distorted V state is easy to obtain in the high magnetic field limit as
described in Sec.~\ref{sec:BEC-hz}.

\subsection{Orthogonal orientation: ${\bf D} \perp {\bf h} = h {\bf \hat{x}}$ }
\label{sec:DMx}
\subsubsection{Classical Ground States}

We will now extend the analysis of the previous section to the case where the magnetic field is applied in the plane of the triangular layers, 
along the $x$-direction for concreteness.  This field orientation is complicated by the fact that the quantum Hamiltonian including DM 
coupling now lacks U(1) spin symmetry; instead the Hamiltonian is symmetric only under discrete spin rotations 
$S^x_{\bf r} \rightarrow S^x_{-{\bf r}}$ and $S^{y,z}_{\bf r} \rightarrow -S^{y,z}_{-{\bf r}}$.  As above, we start by deducing the order 
favored by DM coupling in this field orientation using first-order perturbation theory.  Using Eq.\ (\ref{Salpha}) to parametrize the 
classical ground states along with the constraints in Eqs.\ (\ref{groundstateparameterization}), the first-order correction to the 
energy arising from DM coupling can be written
\begin{eqnarray}
  \langle H_{\rm DM} \rangle &=& -\frac{2DN}{9J}[h(\sin\theta_B \sin\phi_B - \sin\theta_A \sin\phi_A) 
  \nonumber \\
  &+& 9J \sin\theta_A\sin\theta_B\sin(\phi_A-\phi_B)],
  \label{DMenergy2}
\end{eqnarray}
where the angles are subject to the additional constraint
\begin{eqnarray}
  \frac{1}{2}\left[1+\left(\frac{h}{3J}\right)^2\right] &=& \frac{h}{3J}(\sin\theta_A \cos\phi_A + \sin\theta_B \cos\phi_B) 
  \nonumber \\
  &-& \sin\theta_A \sin\theta_B \cos(\phi_A-\phi_B)
  \nonumber \\
  &-&  \cos\theta_A \cos\theta_B .
  \label{constraint}
\end{eqnarray}
One can use Eq.\ (\ref{constraint}) to eliminate $\theta_B$ from Eq.\ (\ref{DMenergy2}) and then minimize the DM energy over the 
remaining angles $\theta_A, \phi_A$, and $\phi_B$.  In this way we find that DM coupling selects coplanar inverted-Y 
states of Fig.\ \ref{ClassicalGroundStates}(e), with 
\begin{eqnarray}
  \theta_{A/B/C} &=& \pi/2
  \nonumber \\
  \phi_A &=& -\phi_B = \cos^{-1}\left[\frac{1}{2}\left(\frac{h}{3J}-1\right)\right]
  \\
  \phi_C &=& 0.
  \nonumber
\end{eqnarray} 

The emergence of planar ground states for this field orientation is quite natural, again because DM interactions favor spins orienting within the triangular lattice plane.  Among the various planar arrangements, inverted-Y states most effectively gain DM energy by keeping neighboring spins as far from parallel as possible (within the constraints set by the ground state manifold).  In contrast, the UUD and V states favored by quantum fluctuations yield a vanishing first-order DM energy because in both configurations spins on two sublattices are parallel.  We note that the favorability of inverted-Y over Y states at low fields is less obvious; the latter also gains DM energy, though to a lesser extent than the former.  

Just as in the ${\bf h} = h{\bf\hat{z}}$ field orientation, incorporating thermal or quantum fluctuations therefore leads to a delicate competition with DM interactions.  The discussion from Sec.\ \ref{ThermalFluctuationsDM} regarding entropic effects applies in this field orientation as well.  Perhaps the most important qualitative distinction with the ${\bf h} = h{\bf\hat{z}}$ field orientation is the additional loss of U(1) symmetry here due to the orthogonality
between ${\bf h}$ and ${\bf D}$ vectors.  This implies the demise of the 1/3 magnetization plateau since the UUD and Y states then no longer constitute distinct phases separated by a spontaneous U(1) symmetry breaking.  
Sec.~\ref{sec:AnisotropicPlusDMquantum} discusses this effect in greater detail; see Fig.\ \ref{fig:M+DM}
for an illustration of the rounding-off of the magnetization plateau by DM interactions.

In the following subsection we will resolve the outcome of the competition between quantum fluctuations and DM interactions at zero temperature, which is qualitatively different from the previously considered ${\bf h} = h{\bf\hat{z}}$ case.

\subsubsection{Quantum Ground States}
\label{sec:DMx_quantum}

The quantum zero-temperature phase diagram in this field orientation is obtained as described in Sec.\ \ref{sec:DMz_quantum}, again treating quantum corrections and DM coupling to first order with $D/J=0.05$ and $S$ extrapolated to $1/2$.  Figure \ref{fig:quantum+dm}(b) depicts our results.
Quantum effects and DM coupling compete in a subtle manner for this field orientation.  The former dominates at low fields, where Y states prevail, while the latter dominates at high fields where inverted-Y states appear. Distorted V order, which reflects a nontrivial compromise between the two competing effects, appears at intermediate magnetization, $1/3 \lesssim h/h_{\rm sat} \lesssim 0.65$.    Note that all states are either perfectly planar or very nearly so here, which again is rather natural. Also notable is the absence of the collinear UUD state in the quantum phase diagram.
As argued above, this is caused by the loss of U(1) spin symmetry in the ${\bf D} \perp {\bf h}$ geometry.

Observe that the quantum phase diagrams predicted for the two field orientations, ${\bf h} \propto {\bf \hat{z}}$ and
${\bf h} \propto {\bf \hat{x}}$, are very different.  At either low or high fields, this implies that at least one phase transition separating 
the umbrella and Y/inverted-Y states ought to appear as one rotates the field from the $z$-direction to the $x$-direction.  
Recent experiments\cite{takano_private} do indeed find a very complicated evolution of the phase diagram as the field rotates in this plane, featuring several intervening phase transitions.  Generalizing the results of this paper to explore this crossover would be an interesting future research direction.

\section{BEC analysis near saturation}
\label{sec:bec}

We will now shift our focus to phases realized at high magnetic fields near saturation, which 
can be elegantly described via Bose-Einstein condensation of magnon excitations
above the fully polarized (saturated) state \cite{batyev1984,nikuni1995,veillette2005,olegt2008,totsuka2009}. 
In this way magnetically ordered phases just below saturation $(h = h_{\rm sat} - \delta h)$ can be understood as a BEC instability of the ground state just above saturation $(h = h_{\rm sat} + \delta h)$.  This limit has the advantage of allowing one to incorporate all of the competing effects of interest---quantum fluctuations, DM coupling, and spatial anisotropy---in a single controlled and coherent framework.

\subsection{Preliminaries: $J = J', D = 0$}
\label{BECisotropic}

At $h > h_{\rm sat}$ the ground state of the Heisenberg Hamiltonian is known exactly: it is 
given by the fully polarized eigenstate of the Hamiltonian in which all spins point `up'.
The excitations are magnons, \emph{i.e.}, plane-wave states of overturned spins.
These are conveniently described by Holstein-Primakoff bosons, similar to Eqs.\ \eqref{eq:HP}.
Since all spins point in the same direction, it suffices to consider only a single boson species here; thus we write
\be
S^z({\bf r}) = S-a^\dagger({\bf r})a({\bf r}) , ~S^+({\bf r}) = \sqrt{2S} ~a({\bf r}) ,
\label{eq:HP2}
\ee
for all ${\bf r}$.  The isotropic Heisenberg Hamiltonian without DM coupling then reads, neglecting $1/S$ and smaller contributions,
\bea
H &=& H_0 + V 
\nonumber \\
  H_0 &=& \sum_{\bf k} S [J({\bf k}) - J({\bf Q}) - \mu] a^\dagger_{\bf k} a_{\bf k}
  \nonumber\\
V &=& \frac{1}{4N} \sum_{\bf k, k', q} [2J({\bf q}) - J({\bf k}) - J({\bf k'} - {\bf q})]\nonumber\\
&& \times a^\dagger_{\bf k+q} a^\dagger_{\bf k'-q} a_{\bf k'} a_{\bf k} .
\label{eq:bec1}
\eea
Here 
\bea
J({\bf k}) &=& \sum_{j = 1}^3 J_{{\bf r}, {\bf r} + {\bm \delta}_j} \cos({\bf k}\cdot{\bm \delta}_j)  \nonumber\\ &=&
 J \left[\cos(k_x) +  2\cos(k_x/2)\cos(\sqrt{3}k_y/2)\right]
\label{eq:bec_J}
\eea
is the Fourier transform of the exchange interaction, and 
\begin{equation}
  \mu = (h_{\rm sat} - h)/S
\end{equation}
is the chemical potential
which controls the state of the bosonic system. For $h > h_{\rm sat}$ the chemical potential is negative
and the ground state is the boson vacuum $|0\rangle$ satisfying $a_{\bf k} |0\rangle = 0$ for all ${\bf k}$.
The wavevector ${\bf Q}$ in Eqs.\ (\ref{eq:bec1}) corresponds to the minimum of the magnon energy and follows from $J({\bf Q}) = \text{min}_{\bf k} [J({\bf k})]$.
A peculiarity of the triangular antiferromagnet \cite{nikuni1995}, and indeed many other frustrated spin systems \cite{totsuka2009},
is the fact that there are two distinct wavevectors minimizing $J({\bf k})$. For the isotropic
two-dimensional triangular antiferromagnet these are $\pm {\bf Q} = (\pm 4\pi/3, 0)$. 
Hence, at $h = h_{\rm sat}$ (or equivalently $\mu = 0$), the magnon dispersion touches zero simultaneously at $\pm {\bf Q}$.  One then needs to understand whether it is energetically favorable to condense magnons at one or both of these wavevectors, \emph{i.e.}, form a single-$Q$ or double-$Q$ condensate.

The analysis proceeds by parametrizing the magnon
modes as follows,
\be
a_{\bf k} = \sqrt{N} \psi_{+Q} \delta_{{\bf k},{\bf Q}} + \sqrt{N} \psi_{-Q} \delta_{{\bf k},- {\bf Q}} + \tilde{a}_{\bf k} .
\label{eq:bec2}
\ee
Here $\tilde{a}_{\bf k}$ describes non-condensed magnons with ${\bf k} \neq \pm {\bf Q}$.
The leading contribution to the energy (per site) is obtained by neglecting the non-condensed particles altogether.
Written in terms of condensate densities $\rho_{1,2} = |\psi_{\pm Q}|^2$, the energy reads
\bea
\frac{E}{N} = - S\mu (\rho_1 + \rho_2) + \frac{1}{2} \Gamma_1 (\rho_1^2 + \rho_2^2) + \Gamma_2 \rho_1 \rho_2 .
\label{eq:bec3}
\eea
The coefficients $\Gamma_{1,2}$ will be defined momentarily below, but first it is useful to discuss the important physics described by this simple equation in some generality.  When $\mu < 0$, the energy is minimized by a vacuum
state with no condensate, $\rho_{1,2} =0$.  Increasing $\mu$ (by lowering $h$) to positive values leads to a condensed state.  
A double-$Q$ condensate emerges for $\Gamma_1 > \Gamma_2$, when
$\rho_1 = \rho_2 = S\mu/(\Gamma_1 + \Gamma_2)$. The physical meaning of this is found
by expressing the spin operators in terms of boson condensates. 
Using $\psi_{\pm Q} = \sqrt{\rho_1} e^{i \theta_{1,2}}$ and writing $\theta_{1,2} = \theta_+ \pm \theta_{-}$,
we find
\bea
S^+_{\bf r} &=& 2 S \sqrt{\frac{2\mu}{\Gamma_1 + \Gamma_2}} e^{i \theta_+} \cos[{\bf Q}\cdot {\bf r} + \theta_{-}] ,
\nonumber\\
S^z_{\bf r} &=&  S - \frac{8\mu S}{\Gamma_1 + \Gamma_2} \cos^2[{\bf Q}\cdot {\bf r} + \theta_{-}].
\label{eq:bec4}
\eea
Hence, the double-$Q$ condensate describes a coplanar magnetically ordered state. Observe that the
$S^z$ spin component is modulated with wavevector $2 {\bf Q}$.
The coplanar state described by Eq.\ \eqref{eq:bec4} thus represents a `supersolid' phase of the magnet. 
A single-$Q$ condensate arises when $\Gamma_1 < \Gamma_2$; in this case $\rho_1 = S\mu/\Gamma_1$ and $\rho_2 =0$ or vice versa. This state 
corresponds to a conventional umbrella (or cone) magnetic structure:
\bea
S^+_{\bf r} &=& S \sqrt{\frac{2\mu}{\Gamma_1}} \exp[i{\bf Q}\cdot {\bf r} + i\theta_1] ,
\nonumber\\
S^z_{\bf r} &=&  S  - \frac{\mu S}{\Gamma_1}.
\label{eq:bec5}
\eea

The classical (order $1/S^0$) expression for $\Gamma_{1,2}$ is given by
\be
\Gamma_1 = J(0) - J({\bf Q}) , ~\Gamma_2 = J(0) - 2J({\bf Q}) + J(2{\bf Q}) .
\label{eq:bec6}
\ee
The accidental degeneracy of the classical isotropic triangular antiferromagnet discussed in Sec.\ \ref{sec:iso_T0}
manifests itself via the relation $\Gamma_1 - \Gamma_2 = J({\bf Q}) - J(2{\bf Q}) = 0$ for 
${\bf Q} = (4\pi/3,0)$.  Thus the coplanar and cone states are degenerate classically.  
At first order in $1/S$ one finds
\be
\Gamma_1 > \Gamma_2
\label{eq:bec7}
\ee 
so that quantum fluctuations select coplanar state, consistent with our earlier spin-wave analysis.  
The corresponding calculation
is sketched in Refs.\ \onlinecite{nikuni1995,totsuka2009} and will not be discussed here for brevity.  Instead, here we take Eq.\ \eqref{eq:bec6} as given and ask how the addition
of weak DM interaction in Eq.\ \eqref{eq:DM} as well as spatial anisotropy influences the delicate balance between coplanar
and umbrella states at high magnetic fields. 

\subsection{Spatially Isotropic Model with \DM~Interactions: $J = J', D \neq 0, {\bf h} = h {\bf \hat{z}}$}
\label{sec:BEC-hz}

In the geometry with ${\bf h} = h{\bf \hat{z}}$ the DM Hamiltonian in Eq.\ \eqref{eq:DM} reduces to a simple quadratic form of magnon operators,
\bea
H_{\rm{DM}} &=& \frac{i D}{2} \sum_{{\bf r}, j = 1,3} 
\Big(S^-_{\bf r} S^+_{{\bf r} + {\bm \delta}_j} - \text{h.c.}\Big)  \nonumber\\
&=& i D S \sum_{{\bf r}, j = 1,3} \Big(a^\dagger_{\bf r} a_{{\bf r} + {\bm \delta}_j} - \text{h.c.}\Big) 
\label{eq:bec8} \\
&=& - 2 N D S[\sin({\bf Q} \cdot {\bm \delta}_1) + \sin({\bf Q} \cdot {\bm \delta}_3)] ( \rho_1 - \rho_2).
\nonumber
\eea
As expected physically, DM coupling breaks the symmetry between $\pm {\bf Q}$ points, and lowers the energy of the $\psi_{+Q}$
condensate (for positive $D > 0$). The ground state follows from minimizing
\bea
\frac{E}{N} & = & - S\mu (\rho_1 + \rho_2) - S g ( \rho_1 - \rho_2) + \nonumber\\
&& + \frac{1}{2} \Gamma_1 (\rho_1^2 + \rho_2^2) + \Gamma_2 \rho_1 \rho_2 ,
\label{eq:bec9}
\eea
where we introduced $g = 2 \sqrt{3} D > 0 $. Following Ref.\ \onlinecite{aharony1975}, we parameterize the densities as
\begin{eqnarray}
   \rho_1 &=& \rho \cos^2\phi
   \nonumber \\
   \rho_2 &=& \rho \sin^2\phi
   \label{rhoparametrization}
\end{eqnarray} and minimize the energy with respect to $\phi$ and $\rho$.

Three possible solutions exist:
\bea
&& (\text{i}) ~\phi = 0, \rho = \frac{S(\mu + g)}{\Gamma_1}, \frac{E}{N} = - \frac{S^2(\mu + g)^2}{2\Gamma_1} \\
&& (\text{ii}) ~\phi= \frac{\pi}{2}, \rho = \frac{S(\mu - g)}{\Gamma_1}, \frac{E}{N} = - \frac{S^2(\mu - g)^2}{2\Gamma_1} \\
&& (\text{iii}) \cos^2\phi = \frac{1}{2} \left[1 + \frac{g (\Gamma_1 + \Gamma_2)}{\mu (\Gamma_1 - \Gamma_2)}\right], 
\rho = \frac{2 \mu S}{\Gamma_1 + \Gamma_2}, \nonumber\\
&& ~~~~~~~~~ \frac{E}{N} = - \frac{\mu^2 S^2}{\Gamma_1 + \Gamma_2} - \frac{g^2 S^2}{\Gamma_1 - \Gamma_2}.
\eea
Solutions (i) and (ii) describe umbrella states formed from a single-$Q$ condensate. For positive $g$ the umbrella corresponding to (i) with $\rho_1 \neq 0, \rho_2 = 0$ has lower energy. The `mixed' solution (iii), which only exists when 
$-1 \leq g (\Gamma_1 + \Gamma_2)/\mu (\Gamma_1 - \Gamma_2) \leq 1$, yields the lowest energy provided
that Eq.\ \eqref{eq:bec7} is satisfied. This solution represents the distorted V state of Fig.\ \ref{ClassicalGroundStates}(d),
discussed in Sec.\ \ref{sec:DMz_quantum} above.  The condensates here satisfy
\bea
|\psi_{+Q}|^2 &=& \frac{\mu S}{\Gamma_1 + \Gamma_2} + \frac{g S}{\Gamma_1 - \Gamma_2}, \nonumber\\
|\psi_{-Q}|^2 &=& \frac{\mu S}{\Gamma_1 + \Gamma_2} - \frac{g S}{\Gamma_1 - \Gamma_2},
\label{eq:bec10}
\eea
and, as a result, the spin structure is non-coplanar,
\bea
S^+_{\bf r} &=& \sqrt{2S} e^{i\theta_{+}}\Big[ (|\psi_{+Q}| + |\psi_{-Q}|)  \cos({\bf Q}\cdot{\bf r} + \theta_{-}) \nonumber\\
&&+ i (|\psi_{+Q}| - |\psi_{-Q}|)  \sin({\bf Q}\cdot{\bf r} + \theta_{-})\Big].
\label{eq:bec11}
\eea
The degree of non-coplanarity is controlled by the DM coupling $g$. The distorted V state has finite longitudinal
chirality [recall Eq.\ \eqref{eq:chirality}] which is proportional to the density imbalance between the $\pm {\bf Q}$ condensates,
\be
\kappa_z \sim 2S (|\psi_{+Q}|^2 - |\psi_{-Q}|^2).
\label{eq:bec12}
\ee

\begin{figure}
\includegraphics[width=3in]{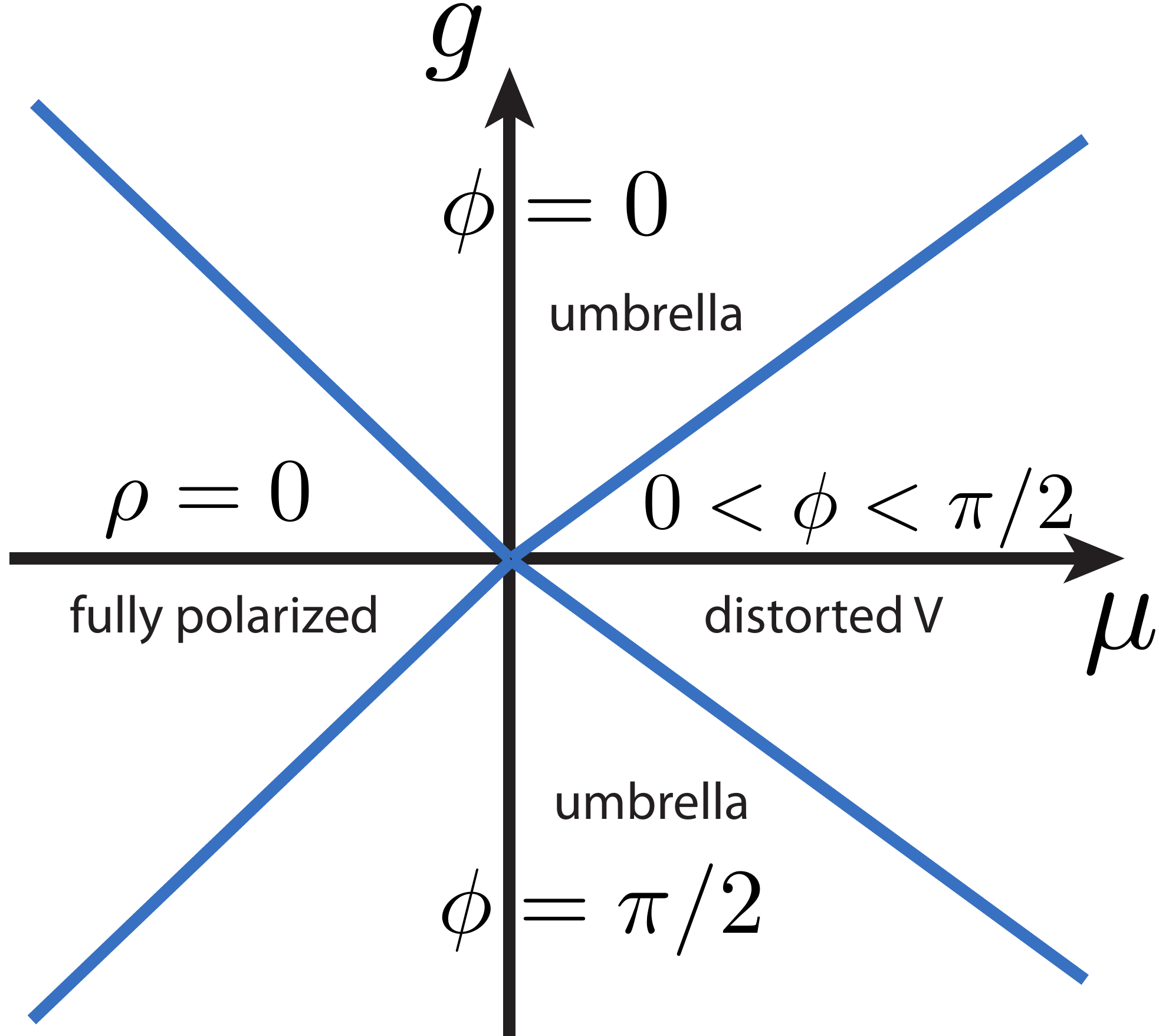}
\caption{(Color online) Phase diagram of the model in Eq.\ \eqref{eq:bec9}. The fully polarized state corresponds to $\rho = 0$, while
all other states exhibit a finite boson density, $\rho > 0$. Umbrella states correspond to $\phi = 0$ and $\pi/2$,
while the mixed state with $0 < \phi < \pi/2$ represents a distorted V phase. Observe that the origin of the phase diagram represents a tetracritical point.}
\label{fig:bec}
\end{figure}

Thus, when the magnetic field orients along the DM vector, the transition from the fully polarized state proceeds in two steps as summarized in the schematic phase diagram of Fig.\ \ref{fig:bec}.  Initially, at $\mu = - g$, one enters the umbrella state
with ordering wavevector ${\bf Q}$.  This state persists at lower fields until $\mu = g (\Gamma_1 + \Gamma_2)/(\Gamma_1 - \Gamma_2)$, where
it is replaced by a distorted V state with condensates at both $\pm {\bf Q}$ wavevectors.
These findings agree fully with a very different analysis in Sec.\ \ref{sec:DMz}, which was carried out for specific values of $D/J$ and $1/S$.  Our results here show that the two phases obtained at high fields near saturation in fact emerge very generally provided DM coupling is weak and quantum fluctuations can be adequately captured perturbatively.  

\subsection{Spatially Isotropic Model with \DM~Interactions: $J = J', D \neq 0, {\bf h} = h {\bf \hat{x}}$ }

To apply the formalism developed in this section to the case where the field orients perpendicular to the DM vector, 
it is simplest to rotate ${\bf D}$ and ${\bf h}$ such that ${\bf h} = h {\bf \hat{z}}$ and
${\bf D} = D {\bf \hat{x}}$. The DM Hamiltonian then reduces to the following combination of {\em three} magnon
operators,
\bea
H_{\rm{DM}} &=& D \frac{\sqrt{2S}}{2i} \sum_{{\bf r}, j = 1,3} 
 \Big[(a_{\bf r} - a^\dagger_{\bf r}) a^\dagger_{{\bf r} + {\bm \delta}_j} a_{{\bf r} + {\bm \delta}_j}  \nonumber\\
&-&  a^\dagger_{\bf r} a_{\bf r} (a_{{\bf r} + {\bm \delta}_j} - a^\dagger_{{\bf r} + {\bm \delta}_j})\Big].
\label{eq:bec13}
\eea
Focusing on the condensates this expression reduces to
\bea
H_{\rm{DM}} &=& D \frac{\sqrt{2S}}{2i} \sum_{{\bf r}, j = 1,3} 
\Big[\psi^*_{+Q} \psi_{-Q} (\psi_{-Q} - \psi^*_{+Q}) e^{-i 3{\bf Q}\cdot{\bf r}} \nonumber\\
&\times& [e^{-i 2 {\bf Q}\cdot{\bm \delta}_j} - e^{-i  {\bf Q}\cdot{\bm \delta}_j}]
+  \{{\bf Q} \to -{\bf Q}\} \Big].
\label{eq:bec14}
\eea
At this point, commensurability of the three-sublattice spin structure acquires crucial importance, since this implies
$e^{\pm i 3{\bf Q}\cdot{\bf r}} = e^{\pm i 4\pi x} = 1$ for all sites ${\bf r}$ of the triangular lattice.
The end result, obtained by separating the magnitudes and phases of the condensates using Eqs.\ (\ref{rhoparametrization}), is rather compact:
\be
H_{\rm{DM}}  = N 4\sqrt{6S} D \rho^{3/2} \sin\theta_{+} \sin(3\theta_{-}).
\label{eq:bec15}
\ee

Notice that Eq.\ (\ref{eq:bec15}) is independent of $\phi$.  This angle is then determined solely by quantum effects which select coplanar order, the detailed structure of which is set by the angles $\theta_\pm$.  In principle, quantum fluctuations do distinguish the latter angles and favor V ordering near saturation; this selection is, however, very weak and appears only when one includes higher-order terms in Eq.\ (\ref{eq:bec3}) (involving six boson fields)\cite{nikuni1995}.  In contrast DM coupling distinguishes these angles more readily.  One can immediately observe that, for $D > 0$, minimization of Eq.\ \eqref{eq:bec15} requires that
$\theta_{+} = \pm \pi/2$.  Let us choose for concreteness $\theta_+ = +\pi/2$.  By consulting Eq.\ \eqref{eq:bec4}, one finds that the plane in which spins order is orthogonal to the (rotated) DM vector.  That is,
\begin{eqnarray}
  S^x_{\bf r} &=& 0 
  \nonumber \\
  S^y_{\bf r} &=& \sqrt{8 S \rho} \cos({\bf Q}\cdot {\bf r} + \theta_{-}).
\label{eq:bec16}
\end{eqnarray}
At the same time, we need to impose $\sin(3\theta_{-}) = - 1$ given our choice for $\theta_+$.  The different solutions for $\theta_-$ simply describe equivalent magnetic structures connected by lattice translations, so we focus for concreteness on $\theta_- = \pi/2$.  
This choice results in $S^y_{\bf r} = -\sqrt{8 S \rho} \sin({\bf Q}\cdot {\bf r})$. For the triangular lattice the product
\be
{\bf Q}\cdot {\bf r} = \frac{2\pi}{3} \nu ~\text{mod} ~2\pi
\label{eq:bec17}
\ee
takes on three inequivalent values: $0$ ($\nu =0$), $2\pi/3$ ($\nu=1$), and $-2\pi/3$ ($\nu=-1$).
Correspondingly, the ordered $S^y$ spin components take on values which are zero ($\nu =0$),
positive ($\nu = 1$) and negative ($\nu = -1$). This represents the three-sublattice  inverted-Y state found previously
in Section\ \ref{sec:DMx} for a specific value of $D/J$ and $1/S$.  The rather different approach adopted here reveals that the onset of inverted Y order induced by DM coupling near saturation is in fact a very generic conclusion.

\subsection{Interplay Between Spatial Anisotropy and \DM~Interactions: $J \neq J', D \neq 0$}
\label{BECanisotropy}

We will now begin addressing for the first time the influence of spatial anisotropy, which can be incorporated in a particularly simple manner near the saturation field.  Let us first address the classical model without DM coupling.  When $J \neq J'$ the Fourier transform of the exchange interaction in Eq.\ (\ref{eq:bec_J}) is now minimized at generally incommensurate wavevectors $\pm {\bf Q}$ given by
\begin{equation}
  {\bf Q} = 2\cos^{-1}{\left(\frac{-J'}{2J}\right)}{\bf \hat{x}}.
  \label{wavevector}
\end{equation}
Consequently the difference between $\Gamma_{1,2}$ defined in Eqs.\ (\ref{eq:bec6}) is non-zero already at the classical level; to order $(J-J')^2/J^2$, one finds $\Gamma_1-\Gamma_2 = -9(J-J')^2/(2J)$.
It follows from our analysis in Sec.\ \ref{BECisotropic} that since $\Gamma_2 > \Gamma_1$ here, incommensurate umbrella states are stabilized classically below the saturation field, even with arbitrarily weak exchange anisotropy.  

The inclusion of quantum fluctuations changes the situation in an interesting way.  Recalling that $\Gamma_1 > \Gamma_2$ due quantum effects in the isotropic limit, to leading order in $1/S$ and second order in exchange anisotropy we have
\begin{equation}
  \Gamma_1 -\Gamma_2 \approx \frac{\alpha J }{S}-\frac{9(J-J')^2}{2J}
\end{equation}
for some constant $\alpha$.  It is now apparent that the planar ordering favored by quantum fluctuations is in fact stable over a finite range of anisotropy, though due to the shift in ${\bf Q}$ such planar configurations will now be incommensurate.  Only when the anisotropy reaches a critical strength are these states supplanted by umbrella order.  

This analysis can be readily extended to incorporate DM coupling as well.  Consider first the field orientation ${\bf h} = h{\bf \hat{z}}$, where the magnetic field and ${\bf D}$ vector are parallel.  As long as spatial anisotropy is sufficiently weak that $\Gamma_1$ exceeds $\Gamma_2$, the phase diagram shown in Fig.\ \ref{fig:bec} remains qualitatively intact (though the umbrella and distorted V states become incommensurate).  As $|J-J'|$ increases, the incommensurate distorted V state arises over a progressively smaller region of the phase diagram until, when $\Gamma_1 = \Gamma_2$, it is removed entirely.  At larger anisotropy only incommensurate umbrella phases appear just below saturation.  

The case where the magnetic field orients perpendicular to the DM vector is even more straightforward.  In the preceding subsection the effectiveness of DM coupling relied critically on the wavevector ${\bf Q}$ being commensurate; see Eq.\ (\ref{eq:bec14}).  
With incommensurate ${\bf Q}$ the corresponding term sums to zero and thus drops out.  
Thus the DM interaction is effectively gone and we then recover the physics discussed above 
for the anisotropic quantum model without DM interactions.

\section{Spatially Anisotropic Model: $J \neq J', D = 0$}
\label{SpatiallyAnisotropicModel}

\subsection{Classical limit}
\label{sec:anisotropy-classical}

In the remaining sections we will endeavor to address the phase diagram in the presence of spatial anisotropy at arbitrary fields.  We start by treating the case without DM interactions for simplicity.  Arbitrarily weak anisotropy $J - J'$ destroys the accidental classical ground state degeneracy
of the isotropic model in favor of an incommensurate spiral ground state at all fields, similarly to the situation described above near saturation.
The ordering wavevector is in fact given by Eq.\ (\ref{wavevector}) for all values of the magnetic field below saturation; the corresponding spin state
is described by 
\begin{equation}
{\bf S}({\bf r}) =  \sin{\theta}\left[\cos{\left({\bf Q}\cdot{\bf r}\right)}{\bf \hat{x}} + 
\sin{\left({\bf Q}\cdot{\bf r}\right)}{\bf \hat{y}}\right] + \cos\theta {\bf \hat{z}} .
\label{incomspin}
\end{equation}

While a unique classical ground state therefore emerges for all fields at zero temperature, the finite-temperature phase diagram is much more complicated and interesting.
This can be anticipated on physical grounds. Indeed, 
as a general rule thermal fluctuations prefer coplanar over non-coplanar states (see the discussion in Sec.\ \ref{sec:isotropic-states}); thus for sufficiently weak anisotropy one can expect that the entropic gain
from planar order should be able to overcome its classical energy cost over a range of temperatures. This means that
planar states should appear {\em above} some critical temperature $T_{\rm pl}$. 
It is reasonable to expect that $T_{\rm pl} \sim (J-J')^2/J$ since the classical energy gain by umbrella states occurs at second order in $J-J'$. It is also clear that
as $J-J'$ increases, the planar states will be gradually pushed towards higher temperatures and disappear
altogether above some critical value (which is magnetic-field dependent) of the spatial anisotropy.

\begin{figure}
\includegraphics[width=3in]{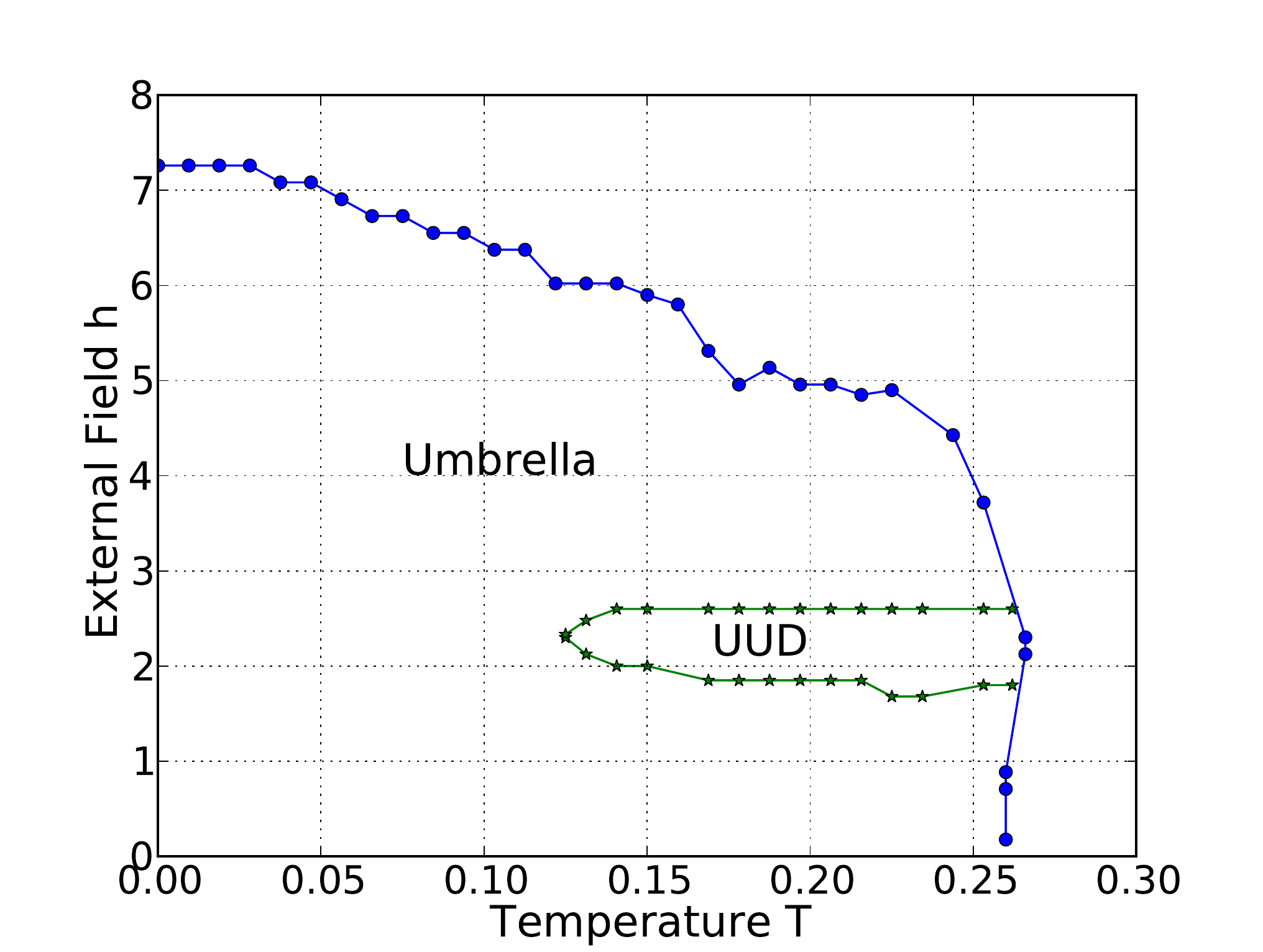}
\caption{(Color online) Phase diagram of the classical anisotropic model with $J'/J = 0.765$.  Interestingly, at this anisotropy strength the entropically favored planar phases are absent (within the resolution of our numerics), save for the collinear UUD state which now requires intermediate temperatures to appear.  }
\label{fig:aniso-diagram}
\end{figure}

Of all the planar phases considered the UUD state, which in fact is collinear,
yields the highest entropy at finite temperature. The relative stability of the UUD state over the Y and V states is clear 
already at the spatially isotropic point: Figure~\ref{classisopd} shows that at finite $T$ the
collinear state expands at the `expense' of the planar ones.
We thus expect the UUD state to persist the most upon deformation of the model's parameters.
Our numerical findings fully support this conclusion and reveal the
rather non-trivial phase structure of the classical model.   Figure \ref{fig:aniso-diagram} displays the 
rough phase diagram obtained using classical Monte Carlo with $J'/J = 0.765$.  This value was chosen
due to its closeness to the estimate for \ccb, and the fact that for this anisotropy strength the wavevector 
${\bf Q}$ `fits' into the $48\times48$ system so that incommensurate orders are not frustrated in the geometries we simulated.  
In these simulations
we have taken $20,000$ Monte Carlo steps for thermalization and $20,000$ more for measurements.

\begin{figure}
\includegraphics[width=3in]{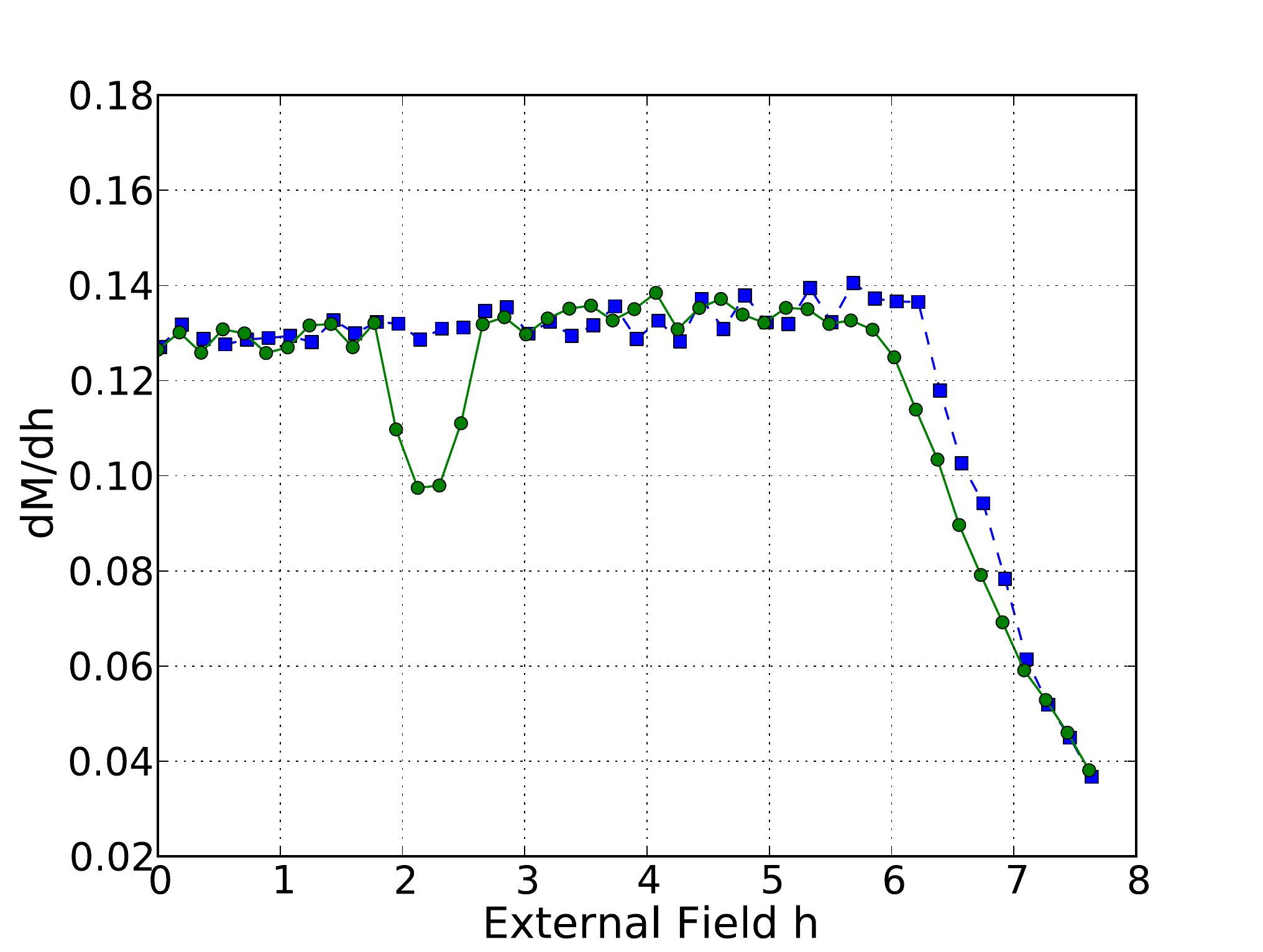}
\caption{(Color online) $dM/dh$ versus $h$ at $T=0.168 J$ (green circles) and $T = 0.12 J$ (blue squares) 
for the classical anisotropic model with $J'/J = 0.765$. }
\label{fig:plateau-aniso}
\end{figure}

\begin{figure}
\includegraphics[width=3in]{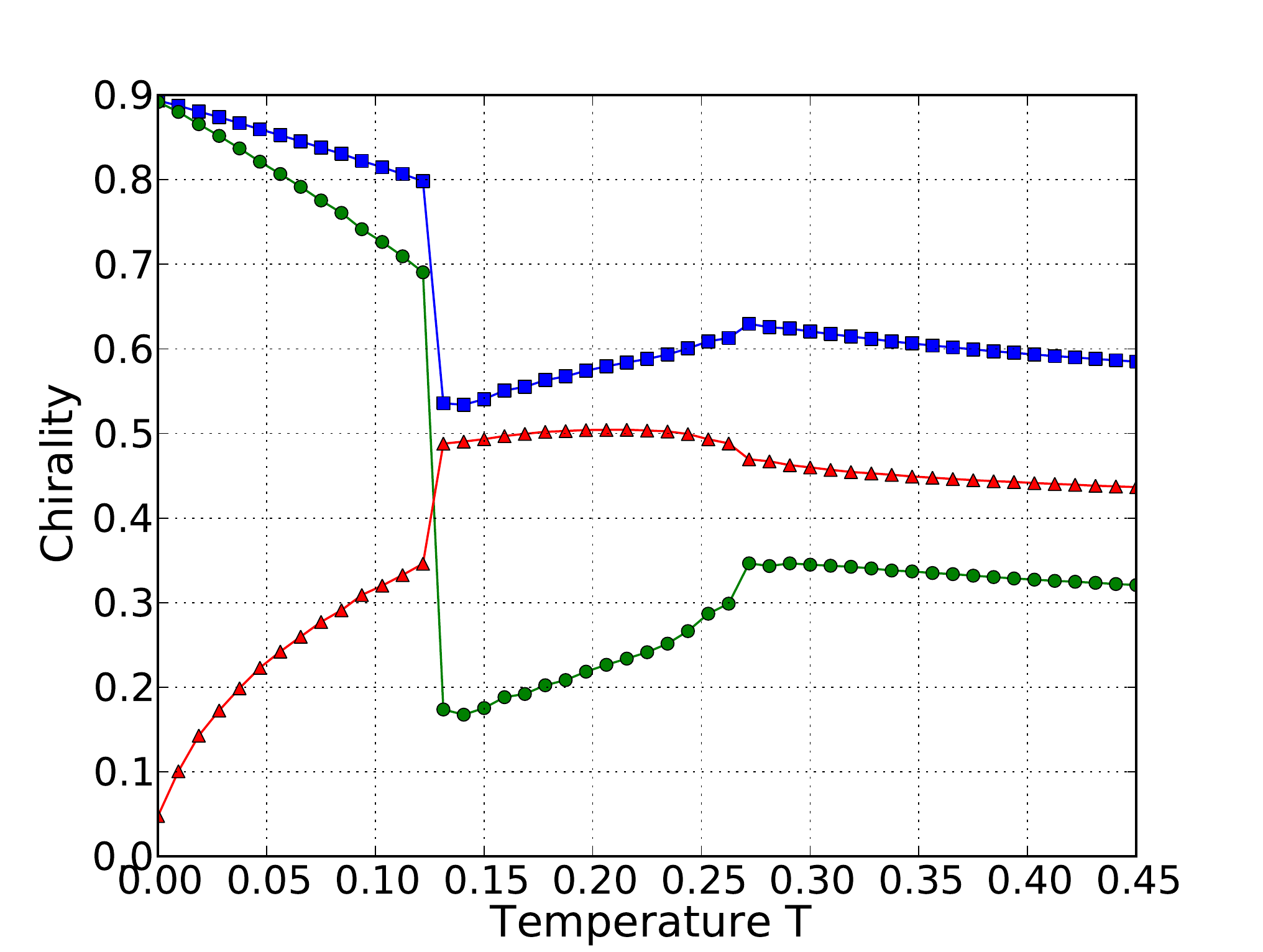}
\caption{(Color online) Chiralities $\kappa$ (blue squares), $\kappa_\perp$ (red triangles) and $\kappa_z$ (green circles)
versus $T$ at $h=2.3 J$ for the classical anisotropic model with $J'/J = 0.765$.  The discontinuous jumps near $T = 0.13 J$ 
signify a first-order transition between umbrella and UUD phases.  }
\label{fig:chirality_Jp_h20}
\end{figure}

Remarkably, the UUD state indeed remains and `floats' above the energetically preferred umbrella phase.  
The UUD order can be clearly identified by the behavior of $dM/dh$ versus field.  As shown in Fig.\ \ref{fig:plateau-aniso}, at an intermediate temperature of $T = 0.168 J$ we observe a pronounced dip in  $dM/dh$, indicating a diminished slope of the magnetization
in the field interval between $h_{c1}$ and $h_{c2}$. Of course slower growth of the magnetization is expected for the magnetization plateau.  
At a lower temperature of $T = 0.12 J$, $dM/dh$ shows no such variation indicating the absence of UUD order.  
This picture is further corroborated by the temperature dependence of the chirality as Fig.\ \ref{fig:chirality_Jp_h20} 
illustrates for $h = 2.3 J$.  One clearly sees a discontinuous jump at $T\approx 0.13 J$ in both the transverse and longitudinal chiralities, 
indicating a first-order finite-temperature transition from the entropy-stabilized UUD state to an umbrella phase.  
Note that, with the possible
exception of the small regions near the UUD boundary
where our numerics do not have enough accuracy to reach any definite 
conclusions, the Y and V planar states are absent in the phase
diagram.  At the anisotropy we analyzed here they are replaced by the energetically favorable umbrella structure.

The already non-trivial phase diagram we obtained here certainly deserves more extensive numerical investigation.  
In particular it would be interesting to perform a systematic study increasing $J-J'$ from zero to explore the collapse 
of the Y/UUD/V states and accompanying onset of umbrella order.  
We note that a similar finite-temperature competition between energetically-favorable and entropically-favorable states
has been previously reported in more complex spin systems: frustrated pyrochlore \cite{pinettes2002,chern2008} and 
Shastry-Sutherland antiferromagnets \cite{moliner09}. It is worth noting that the roots of this behavior 
can be traced to the famous ``Pomeranchuk effect" in $^3$He where the crystal phase of $^3$He, which is characterized
by exponentially weak in distance exchange interaction between localized spins, has higher entropy
than the normal Fermi-liquid phase. 
As a result, upon heating the liquid phase freezes into a solid\cite{pomeranchuk1950,richardson1997}.

\subsection{Pseudo-quantum Ground States}
\label{AnisotropicQuantum}

Our goal now will be to use Monte Carlo numerics to understand the zero-temperature phase diagram when both exchange anisotropy and quantum effects (modeled via the biquadratic approximation discussed in Sec.\ \ref{sec:biquadraticapproximation}) are present.  
Again to avoid becoming trapped in a local energy minimum, the phase diagram was obtained using simulated annealing as described in Sec.\ \ref{sec:biquadraticapproximation}.

\begin{figure}
\includegraphics[width=3.5in]{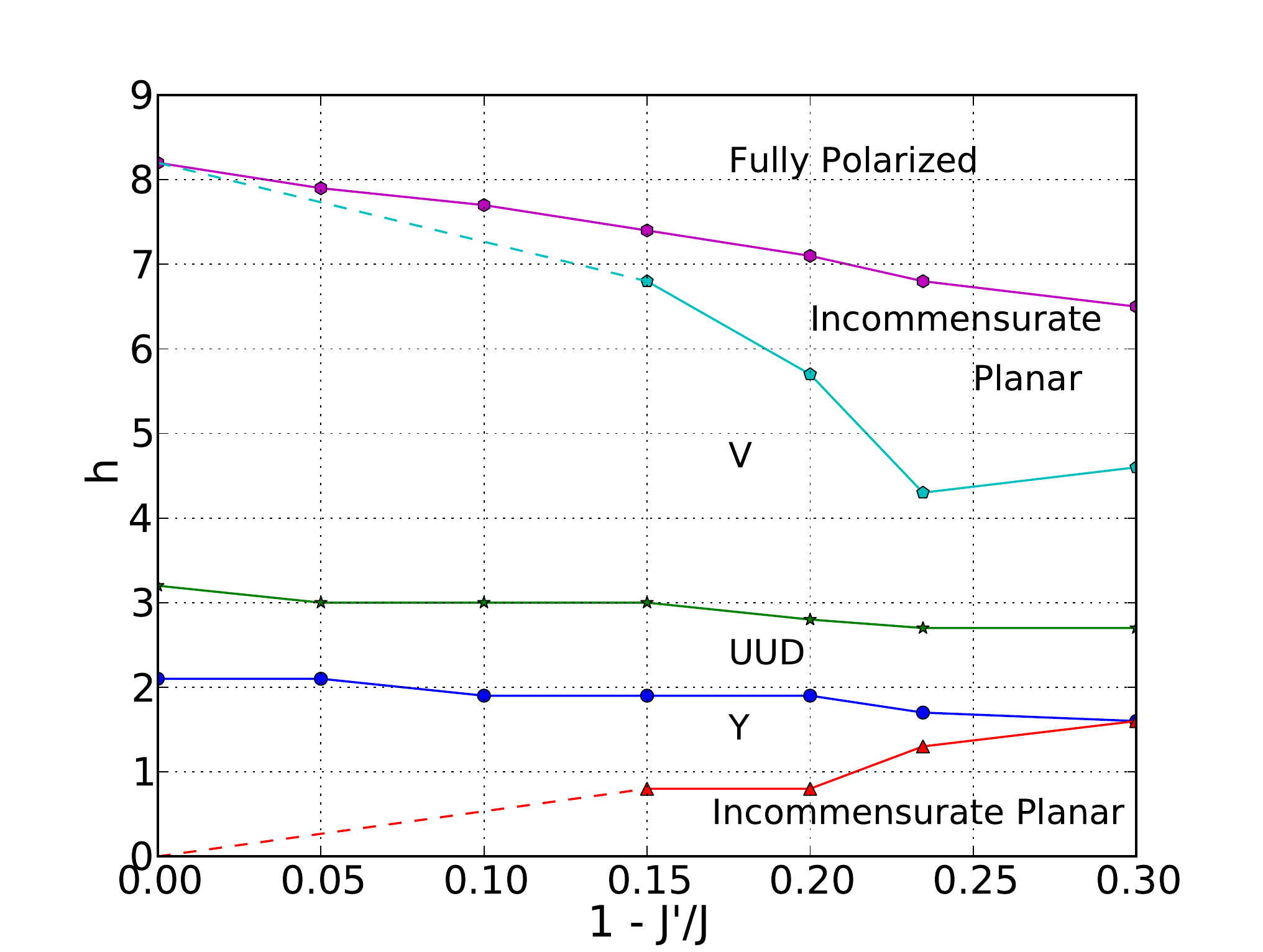}
\caption{(Color online) Ground state phase diagram for the spatially anisotropic triangular antiferromagnet with quantum fluctuations modeled via the biquadratic approximation discussed in Sec.\ \ref{sec:biquadraticapproximation}.  All data points were obtained with Monte Carlo simulated annealing numerics.  Dashed lines represent phase boundaries which are absent due to finite-size effects in our simulations but are expected on general grounds for an infinite system.}
\label{fig:quantumaniso}
\end{figure}

To explore the competition between anisotropy and quantum effects, simulations were performed with fixed $S = 1/2$ but numerous anisotropies $J'/J = 1, 0.95, 0.9, 0.85, 0.8, 0.765, 0.7$.
We used $M, dM/dh, K$ and $\bf{\kappa}$ to construct the phase diagram at each $J'/J$; Fig.\ \ref{fig:quantumaniso} summarizes our findings.  An interesting feature of the phase diagram is the
complete absence of non-coplanar umbrella states, despite the fact that this type of order uniquely minimizes the classical energy at all fields. Even for rather substantial anisotropy $J - J' \sim 0.3J$, quantum
effects (modeled here via the biquadratic interaction) qualitatively alter the magnetization process.

A second interesting feature is the persistence of \emph{commensurate} Y/UUD/V states despite the exchange anisotropy.  The UUD state is particularly stable, and at least within the approximations used is hardly affected by the finite $J-J'$ values studied.  In its vicinity commensurate Y and V states appear over a finite field interval, with the latter being more robust than the former against anisotropy.  All of this is in agreement with the previous large-$S$ analytical investigation of Ref.\ \onlinecite{UUDpaper}.  Incommensurate planar phases, which reflect a nontrivial compromise between quantum fluctuations and anisotropy, appear at low fields and near saturation; in the thermodynamic limit, these are expected to occupy a progressively larger portion of the phase diagram as anisotropy increases.  

It is worth briefly remarking on finite-size effects in our simulations.  First, in Fig.\ \ref{fig:quantumaniso} the high-field incommensurate phase appears over a broader field range at $J'/J = 0.765$ than $J'/J = 0.7$.  This artifact arises because the incommensurate spin structure that would appear in an infinite system is less frustrated by the periodic boundary conditions at the former anisotropy strength.  
Second, we note that incommensurate order sets in at low and high fields only when $J' \lesssim 0.85J$ in our simulations.  Closer to the isotropic limit, they are simply absent.  This reflects finite-size effects arising from the relatively small systems modeled here.  We expect that the phase boundary
between the Y and incommensurate planar states emanates from the $h =0, J' = J$ point.
Similarly, the phase boundary between the V and the incommensurate planar states
extends all the way to $h=h_{\rm sat}, J'=J$; indeed, in Sec.\ \ref{BECanisotropy} we found that arbitrarily weak anisotropy is sufficient to produce incommensurate planar order near saturation.  These phase lines represent commensurate-incommensurate transitions.  We have indicated the expected transitions with dashed lines in Fig.\ \ref{fig:quantumaniso}.

\section{Spatially Anisotropic Model with \DM~Interactions: $J \neq J', D \neq 0$}

\label{AnisotropicPlusDM}

\subsection{Classical Ground States}
\label{AnisotropicPlusDMclassical}

Finally, we are in position to consider the full model featuring all of the ingredients we set out to study: spatial anisotropy, DM coupling, and quantum fluctuations.  As a first step we will establish the phase diagram in the classical case.  For fields ${\bf h} = h{\bf \hat{z}}$ directed along the DM vector the classical ground states simply correspond to incommensurate umbrella order for all fields up to saturation.  This outcome is extremely natural given our earlier findings that for this field orientation both DM coupling and anisotropy separately favor umbrella states.

\begin{figure}
\includegraphics[width=3in]{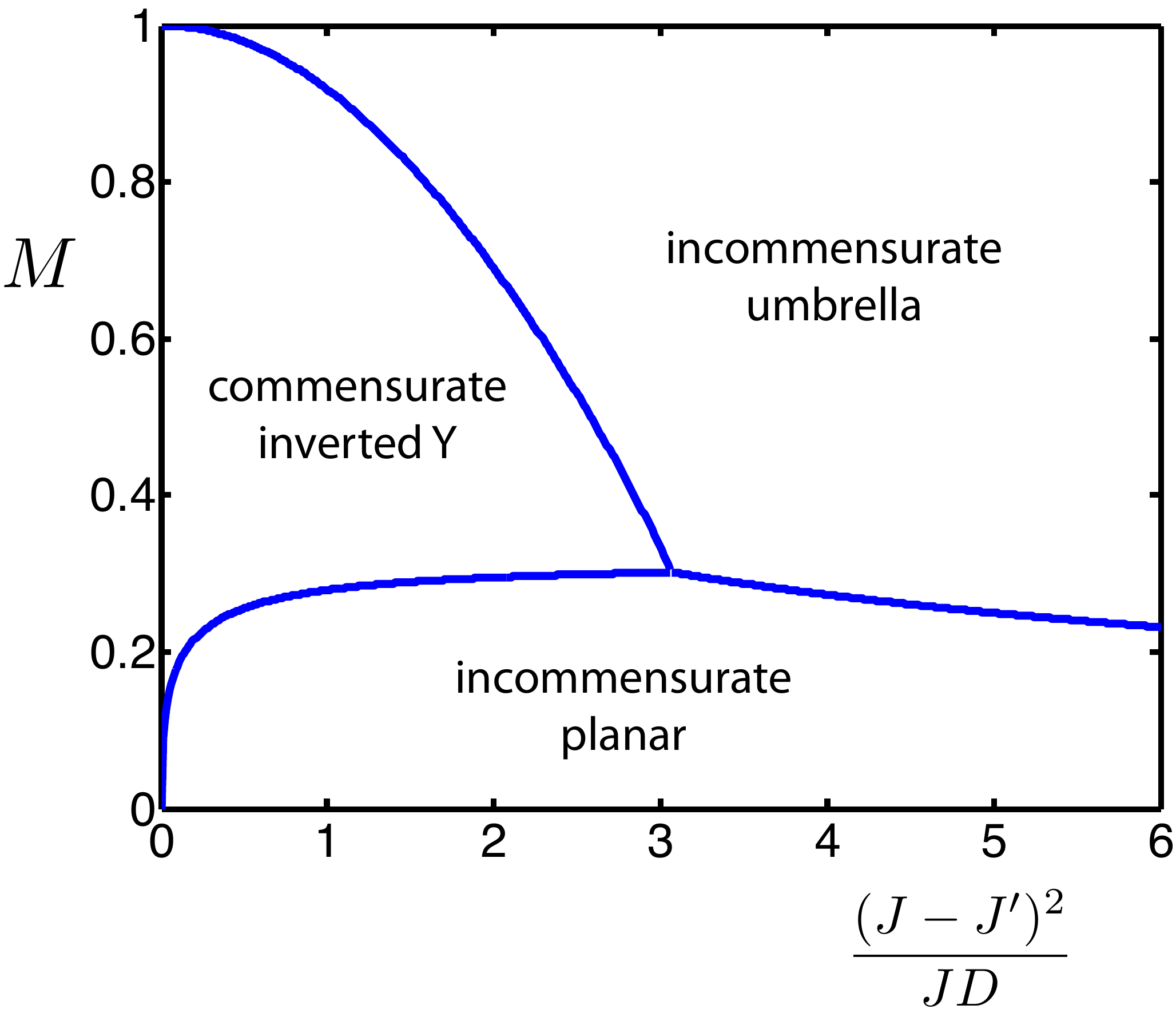}
\caption{(Color online) Classical phase diagram for spins with spatially anisotropic exchange interactions, DM coupling, and a magnetic field applied perpendicular to the DM vector.  Phase boundaries were computed analytically to leading order in DM strength $D/J$ and anisotropy strength $(J-J')^2/J^2$ as described in Appendix \ref{sec:app-aniso+dm}.  }
\label{DManisoclassical}
\end{figure}

The phase diagram is more subtle for fields ${\bf h} = h{\bf \hat{x}}$ oriented perpendicular to the DM vector.  We found earlier that spatial anisotropy favors incommensurate umbrella states, while DM interactions prefer inverted Y order; thus, the resolution of their competition is far from obvious.  Focusing on $J'/J=0.765$ (again, this value minimizes finite-size effects) and $D/J = 0.05$, we find using simulated annealing that
incommensurate coplanar order arises at low fields, $h \lesssim 0.34 h_{\rm sat}$, reflecting a nontrivial compromise between these competing interactions.  
DM coupling dominates at intermediate fields $0.34 h_{\rm sat} \lesssim h \lesssim 0.8 h_{\rm sat}$, where inverted-Y states appear.  The appearance of such a broad commensurate state in the anisotropic system even with only quadratic spin couplings is rather remarkable.  
Finally, at larger fields up to saturation spatial anisotropy dominates, leading to non-coplanar umbrella order. One can in fact analytically estimate the phase boundaries between these three spin states found in our numerics.  The calculation is described in Appendix \ref{sec:app-aniso+dm}, and the resulting classical phase diagram appears in Fig.\ \ref{DManisoclassical}.

\subsection{Pseudo-quantum Ground States}
\label{sec:AnisotropicPlusDMquantum}

Let us now explore how quantum effects---again modeled within the biquadratic approxiation---modify the classical phase diagrams discussed above.  Consider first the ${\bf h} = h{\bf \hat{z}}$ field orientation.  The situation at high fields was already analyzed in Sec.\ \ref{BECanisotropy}, where we found the presence of umbrella order just below saturation, followed by an incommensurate planar state provided anisotropy was not too strong.  At intermediate fields we found in the quantum problem with either DM coupling \emph{or} spatial anisotropy that commensurate planar phases emerged.  It is thus reasonable to anticipate the same outcome when both elements are present, at least for sufficiently weak DM coupling and anisotropy.  At low fields DM coupling led to umbrella order in the quantum problem analyzed in Sec.\ \ref{sec:DMz_quantum}, while the interplay between spatial anisotropy and quantum effects led to an incommensurate planar state in Sec.\ \ref{AnisotropicQuantum}.  

Putting together these findings suggests that the phase diagram for the spatially anisotropic pseudo-quantum model depicted in Fig.\ \ref{fig:quantumaniso} evolves in the following manner as one increases the DM coupling strength from zero.  First, umbrella order immediately begins to `eat away' at the planar phases just below saturation, occupying a progressively larger fraction of the high-field phase diagram as the DM coupling increases.  The low-field incommensurate planar state stabilized by the interplay between spatial anisotropy and quantum fluctuations is more robust against DM coupling.  Only beyond a critical value of the DM coupling does umbrella order begin to take over in the low-field portion of the phase diagram.  Of course more complicated scenarios are all possible, particularly if DM coupling and/or anisotropy are not especially weak; a detailed study of the problem for this field orientation would be interesting to carry out in future work.    

Our main focus, however, is on the low-symmetry field orientation ${\bf h} = h{\bf \hat{x}}$ where the magnetic field and 
DM vectors are orthogonal.  This is the physical situation relevant for the interesting experiments of Ref.\ \onlinecite{Fortune} 
which motivated this study.  We explored the zero-temperature phase diagram here using extensive simulating annealing numerics, 
modeling quantum fluctuations as before using the biquadratic approximation. 
We note that particular care must be taken when performing these simulations to avoid spurious finite-size effects.  In particular, at low fields in $48\times 48$ systems we found an unusual incommensurate planar state exhibiting structure-factor peaks at two incommensurate wavevectors with non-zero momentum $Q_y$ along the $y$-direction.  This phase, however, proved to arise due to finite-size effects---upon increasing the system size to $192\times 192$, order characterized by a single wavevector and vanishing $Q_y$ emerged.

\begin{figure}
\includegraphics[width=3in]{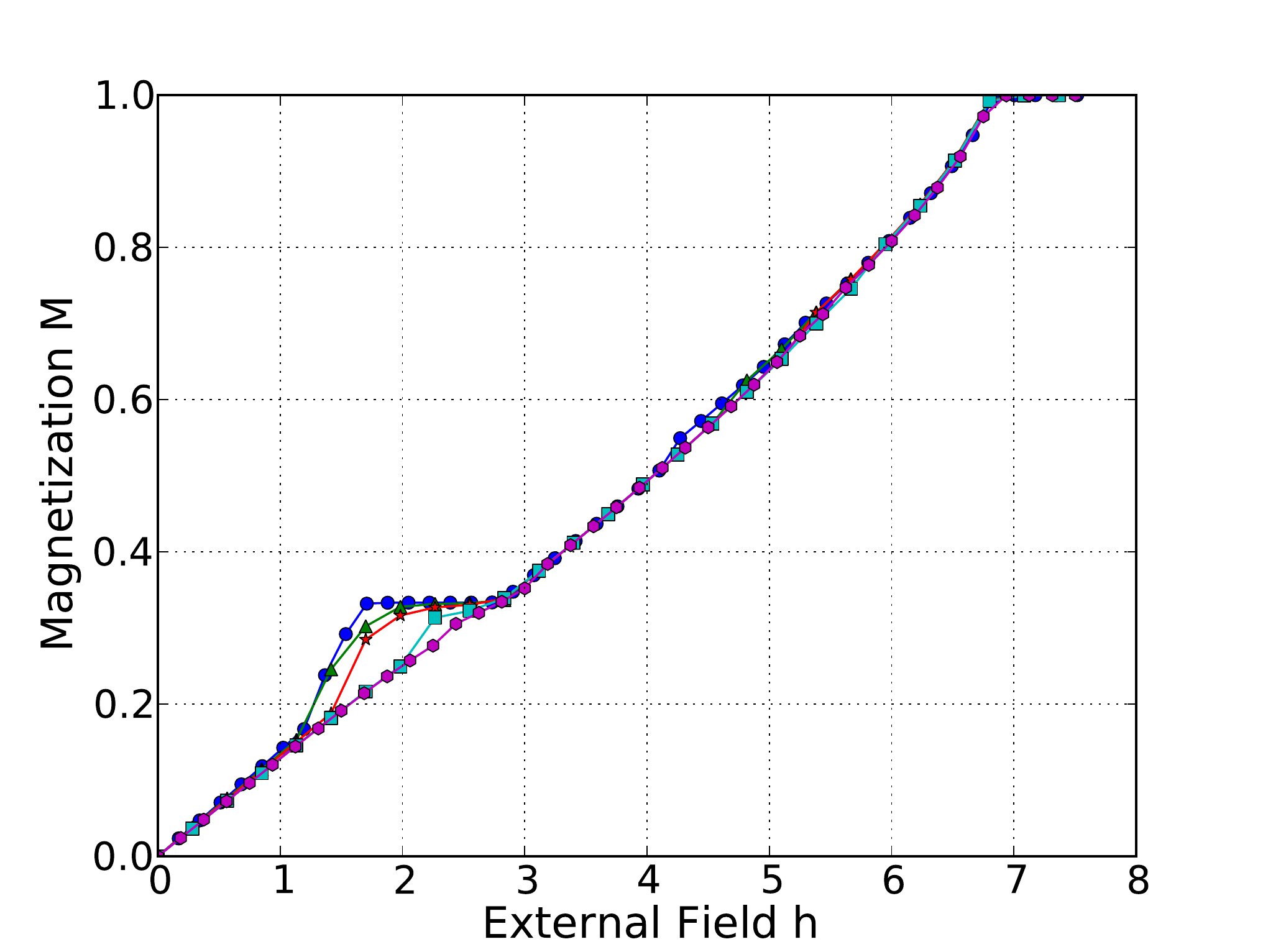}
\caption{(Color online) Magnetization versus field at temperature $T = 0.001J$ for the spatially anisotropic antiferromagnet with $J' = 0.765 J$, DM interactions satisfying ${\bf D} \perp {\bf h}$, and quantum effects modeled via a biquadratic interaction.  The curves shown correspond to different values of the DM strength: $D=0$ (blue circles), $D=0.01 J$ (green triangles), 
$D=0.02 J$ (red stars), $D=0.04 J$ (cyan squares), and $D=0.05 J$ (pink hexagons).  The quick rounding of the lower edge of the plateau reflects the symmetry equivalence of the Y and UUD states when DM interactions are present in this field orientation.}
\label{fig:M+DM}
\end{figure}

\begin{figure}
\includegraphics[width=3in]{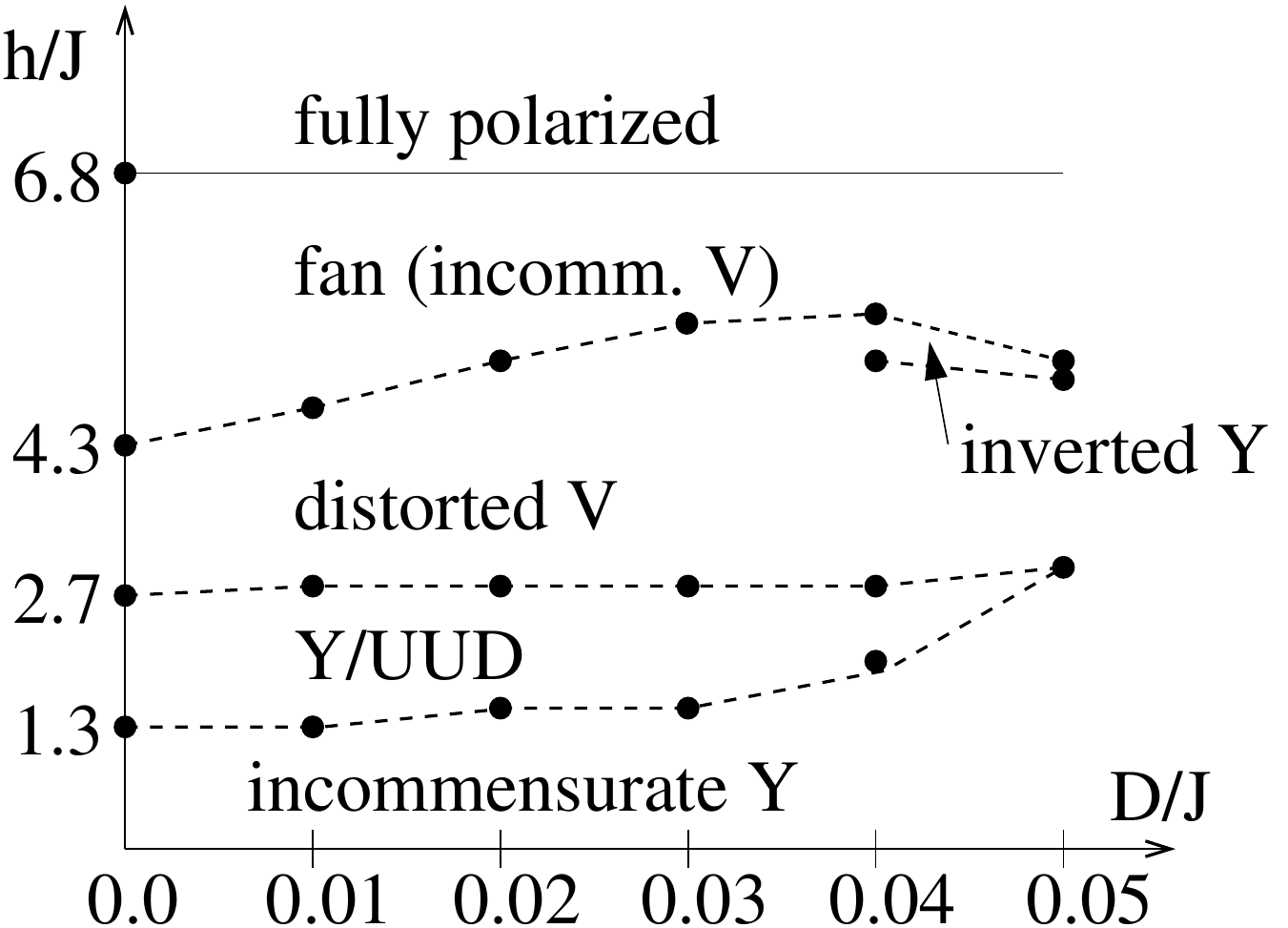}
\caption{Sketch of the phase diagram for the quantum spatially anisotropic model with varying DM coupling strength, $J'/J = 0.765$, and a magnetic field ${\bf h}$ oriented perpendicular to the DM vector.  Data points indicate phase boundaries determined using simulating annealing numerics, with quantum fluctuations modeled via a biquadratic interaction.  
Dashed lines interpolate between
these data points and are drawn for convenience. The location of the inverted Y phase is approximate
as discussed in the main text.}
\label{fig:main_diag}
\end{figure}

Figure \ref{fig:main_diag} summarizes our results. 
As noted in Sec.\ \ref{ClassicalGroundStateSymmetries}, the Hamiltonian
no longer possesses any continuous symmetries; hence some of the phase
boundaries discussed earlier disappear from the phase diagram and become crossovers. One of these is the transition between the Y and
the UUD states.  The difference between these phases in the problem without DM coupling originates from
the finite superfluid component of the Y state, present due to spontaneous breaking
of U(1) spin rotations about the field axis, along with rotation symmetry which the Y state breaks but the UUD state does not. In the ${\bf D} \perp {\bf h}$ problem these
two states are symmetry equivalent as they only break the same discrete lattice symmetries, and are thus not distinct phases.
This shows up in our numerical simulations as a quick rounding of the lower end of the 
magnetization plateau upon increasing the DM coupling at fixed $J'/J$, as illustrated in Fig.\ \ref{fig:M+DM}. 
Simultaneously, for $D$ as small as $D=0.01 J$ the coplanarity $K$ becomes finite, although small,
inside the former UUD interval between the Y and V states.

We also observe persistence of the distorted V state 
at intermediate fields above the (former) UUD state. [Note that as discussed in Sec.\ \ref{ClassicalGroundStateSymmetries} the symmetry distinction between the distorted V and UUD orders persists in this field orientation, so that a bona fide phase transition separates these states.]  In fact it appears that the distorted V state
is the most stable of all the commensurate states considered previously. Because the spins in this phase can smoothly adjust to gain DM energy, this state survives even at the strongest DM coupling strength 
$D/J = 0.05$ considered by us.  By contrast, the previously robust UUD state, having lost its symmetry distinction
from the less stable Y state as discussed above, is seen in Fig.\ \ref{fig:M+DM} 
to essentially disappear for such a strong DM coupling. 

Our simulations also reveal that for sufficiently strong DM coupling $D \gtrsim 0.04 J$
a narrow region of the commensurate inverted Y state appears above the distorted V phase.  
This is quite consistent with the phase diagram of Fig.\ \ref{DManisoclassical} for the classical model described in the
previous subsection: being a prominent phase there,
the inverted Y state is also natural in the quantum problem once quantum effects are sufficiently ``weakened''
by spatial anisotropy and DM interactions.  The transition between the inverted Y and 
distorted V states appears continuous in our numerics and hard to pin down precisely,
in part because the overall extent of this phase is rather narrow. For this reason its
phase boundary in Figure \ref{fig:main_diag} is less accurate than for the other phases.

The remaining high- and low-field regions are found to be occupied by the incommensurate 
coplanar states which owe their stability to quantum fluctuations. 
This is particularly clear for the high-field region near saturation 
where an incommensurate analog of the V state wins over the classical umbrella state
only due to $1/S$ interactions between spin waves, as discussed in Sec.\ \ref{BECanisotropy}.

\section{Conclusions}
\label{Conclusions}

We have explored the phase diagram of a spatially anisotropic triangular lattice quantum antiferromagnet subject
to asymmetric DM interactions and an external magnetic field. By treating spatial anisotropy $(J - J')/J$, quantum fluctuations
due to the finite spin value $S$, and DM coupling $D$ as perturbations of the well-understood isotropic classical 
antiferromagnet, we have found a rich variety of behaviors sensitive to the relative strengths of the perturbations considered.
The root of this richness lies in the large accidental degeneracy of the unperturbed model with
$J'=J$, $S=\infty$ and $D=0$, along with the fact that each of these perturbations favors different ordered states 
in the manifold of accidentally degenerate configurations. 

Our main findings are as follows:

1) In agreement with numerous previous studies, for the isotropic model without DM coupling we observe that quantum fluctuations select coplanar Y and V states and a collinear UUD phase out of infinitely many degenerate states available in the classical limit $S \rightarrow \infty$.  Quantum effects most effectively split this degeneracy in the vicinity of one-third magnetization where nearly collinear low-energy states are accessible.  This implies a greater robustness of the quantum-selected states to additional perturbations in these regions of the phase diagram, a trend that is indeed seen throughout our study.  We also established that quantum effects can be semi-quantitatively modeled within a purely \emph{classical} Hamiltonian by incorporating a biquadratic spin interaction with a field-dependent coupling chosen to mimic $1/S$ corrections to the classical ground state energies.  Such a purely classical model has the virtue of allowing conventional Monte Carlo simulations to be employed to ascertain the (approximate) phase diagram when all the competing interactions of interest are present.  

2) Adding DM interactions introduces two new commensurate states into consideration---distorted V and inverted Y orders. 
With fields applied perpendicular to the triangular lattice plane, DM coupling stabilizes umbrella order classically at all fields up to saturation.  Quantum effects, however, stabilize a distorted V state at intermediate fields, bordered on both sides by the classically driven umbrella phase.  For all in-plane fields up to saturation, DM interactions select inverted Y states classically.  Quantum fluctuations are still more effective at modifying the phase diagram in this field orientation, producing Y order at low fields and a distorted V phase at intermediate fields.  Only at high fields does the classically favored inverted Y state appear in the quantum problem.  It is worth pointing out here that the new distorted V and inverted Y states also appear naturally in the theoretically simple limit of high magnetic fields near saturation, as discussed in
Sec.\ \ref{sec:bec}.
 
3) Spatial anisotropy, on the other hand, prefers non-coplanar umbrella (cone) configurations classically.
Even this simple classical model, which has a unique ground state, harbors some surprises.  We find that for sufficiently small
spatial anisotropy an entropic order-by-disorder selection prevails over energetic considerations over a range of temperatures and magnetic fields.  
This results in an abrupt first-order phase transition from an umbrella state into a fluctuation-stabilized
UUD state at finite temperature; see Sec.\ \ref{sec:anisotropy-classical}. This physically reasonable
but so far unexplored feature of the classical Heisenberg model with anisotropic exchange interactions
deserves more extensive numerical investigation on its own. 

4) The competition between spatial anisotropy and quantum effects (simulated within an effective classical model with biquadratic spin couplings) results in a rich phase diagram shown in Fig.\ \ref{fig:quantumaniso}.  These competing ingredients compromise to form incommensurate versions of the Y and V states at low and high fields, respectively.  Remarkably, the intermediate-field region of the phase diagram in Fig.\ \ref{fig:quantumaniso} is only weakly affected by the anisotropy strengths we analyzed, with the commensurate Y, UUD, and V states favored by quantum effects all appearing prominently.   
In particular, the UUD state which underlies the one-third-magnetization plateau, shows no obvious reduction in its width.  
All these features of the phase diagram agree well with previous analytical \cite{UUDpaper} and numerical \cite{tay2010}
studies.

5) When spatial anisotropy, DM interactions, and quantum effects (again incorporated in an effective classical Hamiltonian) are all present, we obtain the phase diagram of Fig.\ \ref{fig:main_diag}, which contains most of the phases reviewed above.
It is interesting to compare this with the experimental phase diagram of Ref.\ \onlinecite{Fortune} which features up to nine 
different phases. Clearly our phase diagram is less diverse but still contains, for $D\approx 0.04 J$, five different
phases, most of which are coplanar. In particular it is tempting to speculate that one of the puzzling
high-field phases (B, III, IV or 2/3, in the notation of Ref.\ \onlinecite{Fortune}) can be identified with the 
inverted Y state appearing near the phase boundary between the distorted V  and the incommensurate planar phases.
Further experiments that directly probe the spin structure in the various phases seen are, however, required to make a more definitive comparison with our results.  

We emphasize here that nowhere in our study did we find hints of a possible two-thirds-magnetization
plateau suggested by several experimental papers \cite{ono2004,ono2005,Fortune}.
In this regard the possibility of this novel magnetization plateau, at present identified theoretically
only once \cite{miyahara2006}, requires more careful investigations.
It must also be kept in mind that our treatment of quantum fluctuations is only perturbative,
and certainly in the $S=1/2$ model at least the phase boundaries will be quantitatively different from what we have found.
Whether or not strong quantum fluctuations can induce entirely new states, unseen in our 
semiclassical analysis, remains an interesting open question. 

More broadly, our analysis relies on the observation that near the spatially isotropic point,
where the relevant low-energy states form a three-sublattice pattern,
the only relevant DM interaction is that given in Eq.\ \eqref{eq:DM}.
This implies that the phase diagram should be insensitive to the orientation of the
magnetic field in the ${\bf b}-{\bf c}$ plane of the material. Recent experimental 
findings support this statement \cite{takano_private}, and, in our view, should
be interpreted as lending strong support to our perturbative approach to \ccb .
It is worth pointing out that this is certainly
not the case for the isostructural material \ccc; there it has been established
experimentally that phase diagrams for the cases ${\bf h}\parallel {\bf b}$ and
${\bf h}\parallel {\bf c}$ are indeed different \cite{tokiwa2006}. That difference has been
attributed to additional symmetry-allowed DM interactions \cite{starykh10}
and, fundamentally, is tied to the fact that \ccc\ is very much a quasi-one-dimensional material, in contrast to \ccb.

Another, and so far little explored, way to gain insight into the rich physics
of frustrated antiferromagnets is provided by a magnetic system's response to various impurities.
\ccb\ has been probed in this way \cite{ono2005}, and initial theoretical investigations
have appeared recently.\cite{wollny2011}  

We hope that our study will stimulate further 
theoretical and experimental studies of \ccb, as well as related materials and models, in the near future.

\begin{acknowledgements}
We would like to acknowledge helpful conversations and discussions with Leon Balents, Andrey Chubukov, Michel Gingras,
Olexei Motrunich, Roderich Moessner, Gil Refael, Yasu Takano, and Mike Zhitomirsky. 
We are grateful to Yasu Takano for sharing unpublished experimental results with us.
We thank the Center for High Performance Computing
at the University of Utah for their generous computer time allocation in the initial stages
of this project.  This research was supported by the National Science Foundation through grants  
DMR-0808842 (S.H.\ and O.A.S.), DMR-1055522 (J.A.) and DGE-0707460 (C.G.), as well as Victoria B.\ Rodgers and the Rose Hills foundation (C.G.).  
\end{acknowledgements}
\appendix
\section{Symmetry-allowed \DM~interactions in Cs$_2$CuBr$_4$}
\label{sec:appDM}

The general form of the DM interaction, consistent with crystal symmetry of \ccb,
was derived in Ref.\ \onlinecite{starykh10} for the isostructural material \ccc.
For the three-dimensional crystal with layers indexed by $z$, it reads
\begin{eqnarray}
H_{\rm{DM}} &=& \sum_{\bf r} ({\bf D'} \cdot {\bf S}_{\bf r} \times {\bf S}_{{\bf r}+{\bm \delta}_2} + 
{\bf D}^+ \cdot {\bf S}_{\bf r} \times {\bf S}_{{\bf r}+{\bm \delta}_1} \nonumber\\ 
&-& {\bf D}^- \cdot {\bf S}_{\bf r} \times {\bf S}_{{\bf r}-{\bm \delta}_3})
\end{eqnarray}
where
\begin{eqnarray}
{\bf D'} &=& 2D_z(-1)^z{\bf \hat{z}} + D_y(-1)^y{\bf \hat{y}} ,\\
{\bf D}^\pm &=& D_z'(-1)^z{\bf \hat{z}} \pm D_x'(-1)^{y+z}{\bf \hat{x}} + D_y'(-1)^{y+z}{\bf \hat{y}} .
\end{eqnarray}
We are, however, interested in a 2D triangular antiferromagnet so will henceforth consider only one layer with $z = 0$ for concreteness.  We now show that for a weakly deformed triangular antiferromagnet, which is an 
appropriate characterization of \ccb, only two DM couplings $D_z$ and $D_z'$ 
need to be retained to leading order in perturbation theory.
Furthermore, $D_z$ and $D_z'$ yield first-order contributions which are in fact identical in nature
and can thus be combined into a single DM coupling oriented along the crystal
$z$-axis. 

These results follow from the observation that all classical ground states of the isotropic model form a three-sublattice structure which repeats in the $y$-direction; see Fig.\ \ref{LatticeFig}. 
 Let us first consider the term $D_y$. Because of the repeating structure in the $y$-direction, 
we see that the cross products of spins from the A, B and C sublattices cancel out due
to the oscillating $(-1)^y$ factor in the $D_y$ term. 
Thus $D_y$ can only appear at second-order or higher in perturbation theory, and for this reason 
can be neglected. Similar reasoning dictates that $D_x'$ and $D_y'$, which also include 
oscillating $(-1)^y$ factors, do not contribute to first order and can thus be omitted as well.

Consider now the remaining terms, $D_z$ and $D_z'$.
From Fig.\ \ref{LatticeFig} we see that ${\bf S}_{{\bf r}+{\bm\delta}_1}$ and ${\bf S}_{{\bf r}-{\bm \delta}_2}$ 
correspond to the same sublattice. 
Likewise, spins ${\bf S}_{{\bf r}+{\bm \delta}_2}$ and ${\bf S}_{{\bf r}-{\bm \delta}_3}$ can also 
be identified. Consequently, the most general DM Hamiltonian above, when evaluated in an arbitrary three-sublattice classical ground state, reduces to the following expression,
\begin{eqnarray}
  H_{\rm{DM}} &\rightarrow& \sum_{\bf r} {\bf \hat{z}}\cdot(2D_z {\bf S}_{\bf r} \times {\bf S}_{{\bf r}+{\bm \delta}_2} + 
D_z'{\bf S}_{\bf r} \times {\bf S}_{{\bf r}+{\bm \delta}_1} \nonumber\\ 
&-& D_z' {\bf S}_{\bf r} \times {\bf S}_{{\bf r}-{\bm \delta}_3})
\nonumber \\
&=& \sum_{\bf r} {\bf \hat{z}}\cdot[(D_z {\bf S}_{\bf r} \times {\bf S}_{{\bf r}+{\bm \delta}_2} - D_z' {\bf S}_{\bf r} \times {\bf S}_{{\bf r}-{\bm \delta}_3})
\nonumber\\ 
&-& (D_z {\bf S}_{{\bf r}} \times {\bf S}_{{\bf r} - {\bm \delta}_2} - D_z'{\bf S}_{\bf r} \times {\bf S}_{{\bf r}+{\bm \delta}_1}) ]
\nonumber \\
&\rightarrow& (D_z-D_z')\sum_{\bf r} {\bf \hat{z}}\cdot({\bf S}_{\bf r} \times {\bf S}_{{\bf r}-{\bm \delta}_3}
-{\bf S}_{\bf r} \times {\bf S}_{{\bf r}+{\bm \delta}_1})
\nonumber \\
&=& (D_z'-D_z)\sum_{\bf r} {\bf \hat{z}}\cdot[{\bf S}_{\bf r} \times({\bf S}_{{\bf r}+{\bm \delta}_1} + {\bf S}_{{\bf r}+{\bm \delta}_3})].
\end{eqnarray}
Hence, defining ${\bf D} = (D_z-D_z'){\bf \hat{z}} \equiv D{\bf \hat{z}}$, we arrive at the DM Hamiltonian in 
Eq.\ \eqref{eq:DM}.

\section{Analytical determination of phase boundaries in Figure \ref{DManisoclassical}}
\label{sec:app-aniso+dm}

In this Appendix we will sketch the derivation of the phase boundaries depicted in Fig.\ \ref{DManisoclassical}.  
We adopt a variational approach and calculate the energies for the three phases found in simulations---incommensurate planar, 
commensurate inverted-Y, and incommensurate umbrella states---to determine which minimizes the energy as a function of field.  
Of course there is no guarantee that only these three phases are relevant, but our numerical findings suggest 
that this is the case for weak anisotropy and DM coupling.  

For concreteness, let us take the field in the $y$-direction.  For the incommensurate planar state, we work at low fields and parametrize the spins as
\begin{equation}
  {\bf S}_{\bf r} = \cos[{\bf Q}\cdot{\bf r} + \phi({\bf r})]{\bf \hat{x}} + \sin[{\bf Q}\cdot{\bf r} + \phi({\bf r})]{\bf \hat{y}},
\end{equation}
where ${\bf Q}$ is the incommensurate wavevector from Eq.\ (\ref{wavevector}).  With $\phi({\bf r}) = 0$, this expression yields the exact classical ground state at zero field and in the absence of DM coupling.  [We will neglect the dependence of ${\bf Q}$ on the DM coupling here, which results in energy corrections of order $(D/J)^2$.]  In finite fields the function $\phi({\bf r})$ becomes non-zero and gives rise to a net magnetization along the $y$-direction; it suffices to take
\begin{equation}
  \phi({\bf r}) = 2M\cos{\bf Q}\cdot {\bf r}.  
\end{equation}
Up to order $h^2$, $M$ represents the magnetization which is given by
\begin{eqnarray}
  M = \frac{h J^3}{(2J+J')[4J^3 + J'^3 -2J^2(J'-D\sqrt{4-(J'/J)^2})]}.
  \nonumber \\  
\end{eqnarray}
(This expression can be obtained by evaluating the energy to second order in $h$, and then minimizing the expression to find $M$.)  The incommensurate planar energy per site, to order $h^2$, is then
\begin{eqnarray}
  \frac{E_{\rm planar}}{N} &=& -\bigg{\{} hM + \frac{D}{J}\sqrt{4-(J'/J)^2} [J-(2J+J')M^2] 
  \nonumber \\
  &+& \frac{2J^2 + J'^2}{2J} - \frac{(2J+J')^2(2J^2-2JJ'+J'^2)M^2}{2J^3}\bigg{\}}.
  \nonumber \\
\end{eqnarray}

Both the inverted-Y and umbrella states can be easily found at arbitrary fields.  For the inverted-Y state, we have
\begin{eqnarray}
  {\bf S}_A &=& {\bf \hat{y}}
  \\
  {\bf S}_B &=& \sqrt{1-\frac{1}{4}(3M-1)^2}{\bf \hat{x}} + \frac{1}{2}(3M-1){\bf \hat{y}}
  \\
  {\bf S}_C &=& -\sqrt{1-\frac{1}{4}(3M-1)^2}{\bf \hat{x}} + \frac{1}{2}(3M-1){\bf \hat{y}},
\end{eqnarray}
which yields an energy per site
\begin{eqnarray}
  \frac{E_{\rm inverted-Y}}{N} &=& -\bigg{[}\frac{1}{2}(J+2J') + hM - \frac{3}{2}(J+2J')M^2 
  \nonumber \\
  &+& D\sqrt{3(1+3M)}(1-M)^{3/2}\bigg{]}.
\end{eqnarray}
The spin configuration for the incommensurate umbrella state is (again ignoring renormalization of ${\bf Q}$ by $D$) given by Eq.\ (\ref{incomspin}); the corresponding energy per site is
\begin{eqnarray}
  \frac{E_{\rm umb}}{N} &=& -\bigg{[}\frac{2J^2 + J'^2}{2J} + hM -\frac{(2J+J')^2}{2J}M^2\bigg{]}.
\end{eqnarray}

To leading order in $D/J$ and $(J-J')^2/J^2$, the incommensurate planar and inverted-Y state energies cross at a magnetization $M_1$ satisfying
\begin{eqnarray}
  \frac{(J-J')^2}{2J}(1-10M_1^2) &=& \sqrt{3}D[(1-M_1)^{3/2}\sqrt{1+3M_1}
  \nonumber \\
  &-&1+3M_1^2].
\end{eqnarray}
For $M<M_1$ the planar state has lower variational energy, while for $M>M_1$ the inverted-Y state wins.  Taking $D/J = 0.05$ and $J'/J = 0.7$, we find $M_1 \approx 0.3$.  Although our calculation of the planar energy was perturbative in the field and is thus most reliable near $M = 0$, this approximate result for $M_1$ agrees remarkably well with the lower phase boundary determined numerically for these parameter values.  
The umbrella state and inverted-Y energies balance at a magnetization $M_2$ which satisfies
\begin{eqnarray}
  \frac{(J-J')^2}{2J}(1+M_2) = D\sqrt{3(1+3M_2)(1-M_2)}.
\end{eqnarray}
In this case the inverted-Y state yields a lower energy for $M<M_2$.  With $J'/J = 0.7$ and $D/J = 0.05$ we obtain $M_2 \approx 0.7$, also in excellent agreement with numerics.  
At a critical anisotropy strength $(J-J')^2/J \approx 3D$, the magnetization values $M_1$ and $M_2$ coincide, and for larger anisotropy the inverted Y state no longer appears in the phase diagram.  We then have only two phases which compete---the incommensurate planar and umbrella states.  To leading order in $D/J$ and $(J-J')^2/J^2$, their energies cross at a magnetization
\begin{eqnarray}
  M_3 = \left[\frac{9(J-J')^2}{2\sqrt{3}JD}+3\right]^{-1/2}.
\end{eqnarray}
For $M>M_3$ the incommensurate umbrella phase emerges, whereas for $M<M_3$ incommensurate planar order appears.  Plotting the phase boundaries $M_{1,2,3}$ derived above leads to the classical phase diagram of Fig.\ \ref{DManisoclassical}.

\end{document}